\documentclass[fleqn,10pt]{wlscirep}
\usepackage[utf8]{inputenc}
\usepackage[T1]{fontenc}
\usepackage{graphicx}
\usepackage[export]{adjustbox}
\usepackage{bm}
\usepackage[sorting=none, style=numeric-comp, maxnames=99, minnames=1]{biblatex}
\usepackage{lineno}
\usepackage{glossaries}
\usepackage{relsize}
\usepackage{csquotes}
\makeglossaries
\newglossaryentry{attention}
{
        name=attention,
        description={A mechanism in transformers that dynamically weighs the pairwise importance of different input elements in a sequence, or their features, based on the attention calculation. The dot-product attention calculation consists of a linear projection of the inputs to the transformer module into keys, queries and values, where the attention matrix is the result of a softmax applied on the dot product between keys and queries scaled by the dimension of the keys. The final attention calculation involves taking the weights obtained from the attention matrix and computing the dot product with the values}
}

\newglossaryentry{self-attention}
{
        name=self-attention,
        description={A specific form of attention that is calculated within the layers of either the encoder or the decoder of a transformer, not between these two modules. This means the inputs projected into keys, queries and values all come from the encoder, or all come from the decoder}
}

\newglossaryentry{multi-head attention}
{
        name=multi-head attention,
        description={A variation on attention where instead of the keys, queries and values used to calculate attention coming from a single linear projection of each of the inputs, the inputs are linearly projected into keys, queries and values $h$ times, where $h$ is the number of heads in the multi-headed attention. This means the attention calculation is calculated in parallel across all $h$ heads, to be concatenated and once again projected, and ideally $h$ different linear projections of the inputs are learned. Multi-head attention can be self-attention or attention between an encoder and decoder}
}

\newglossaryentry{encoder}
{
        name=encoder,
        description={The part of the transformer that processes and represents input data. In the context of genomics, the encoder takes DNA sequences and creates a numeric representation of the features of this sequence, which can later be used by the decoder to make predictions}
}

\newglossaryentry{decoder}
{
        name=decoder,
        description={The segment of the transformer that produces a sequence output, often based on the representation generated by the encoder. In the context of genomics, this decoder can take representations of DNA sequences that were learned by the encoder and reconstruct a desired output sequence}
}

\newglossaryentry{pre-training}
{
        name=pre-training,
        description={Training a model on a large dataset that is not necessarily task-specific, leveraging more data to learn underlying patterns or structures within the data. Later the model can be \textbf{fine-tuned} for specific tasks using a smaller well-curated dataset. \textbf{Unsupervised pre-training} is a form of pre-training whereby a large corpus of unlabelled data is used to pre-train the model, requiring only a small labelled dataset to be fine-tuned. This allows for available but unlabelled data to contribute to model performance, and improve the results the model would have on an otherwise limitingly small dataset},
}

\newglossaryentry{fine-tuning}
{
        name=fine-tuning,
        description={Adjusting a pre-trained model to adapt it for a particular application through training it on a smaller, task-specific dataset. Sometimes this involves ``freezing'' the majority of the model's weights learned in pre-training and only adjusting a smaller number of weights during fine-tuning, as opposed to all trainable weights}
}

\newglossaryentry{CNN}
{
        name=CNN,
        description={A convolutional neural network. This network has a series of convolutional layers built upon interconnected neuron subsets, or convolutions. Here, neurons that are spatially proximate are linked, while those distant are not (see Supplemental Figure \ref{suppfig1} (b)). These nearby neurons share the weight, or connection strength, which is collectively updated during the backpropagation process. The degree of neuronal proximity dictating their connections is controlled by a hyperparameter known as the ``kernel size''. In genomics, these kernels convolve over DNA sequence, represented by a one-hot encoding of the \textbf{ A T G C} nucleotides (sometimes including an ``N'' nucleotide), and learn to recognize specific subsequences or motifs. In DNA, these kernels are 1D convolutions over 4-channels (or 5, if ``N'' is included) representing the four nucleotides. This can be conceptualized similarly to convolutions applied over a coloured image, where 2D kernels convolve over 3-channels (Red Green Blue, or RGB, channels). These kernels compute the sum of elementwise multiplication between the one-hot encoding of DNA and the filter, to create pattern matches, or motifs, which can later be compared to experimentally determined motifs}
}

\newglossaryentry{kernel}
{
        name=kernel,
        description={The filter used in convolutional layers. In the context of genomics, it computes the sum of elementwise multiplication between itself and the one-hot encoding of DNA to create pattern matches, or motifs, which can later be compared to experimentally determined motifs. The greater the value of the sum over the elementwise multiplication between kernel and sequence, the greater the concordance between the kernel and sequence. Kernels with high values in positions that match up to the sequence encoding, and negative values in all other positions (to penalize differences), can capture subsequences very effectively}
}

\newglossaryentry{RNN}
{
        name=RNN,
        description={A recurrent neural network. RNNs are specialized neural networks often equipped with ``memory'' that allows the model to capture temporal or spatial dependencies within the sequence data (see Supplemental Figure \ref{suppfig1} (c)). RNNs include Long Short Term Memory Networks (LSTMs), Bi-directional LSTMs (BLSTMs) and Gated Recurrent Units (GRUs). An RNN operates on data sequentially, where each sequence element is subjected to the same computational process. Each computation requires a new sequence element and the ``memory'' of previous elements, yielding an updated ``memory'' state that is continuously propagated throughout the sequence}
}

\newglossaryentry{DFNN}
{
        name=DFNN,
        description={A deep feed-forward neural network. This is the earliest deep learning model architecture. A fully-connected DFNN has a series of densely connected layers such that every neuron in a given layer maintains connections to all neurons in the succeeding layer. This property allows for precise modulation of the connection strength between each neuron during the backpropagation process, as visualized in Supplemental Figure \ref{suppfig1} (a) }
}
\newglossaryentry{pretext task}
{
        name=pretext,
        description={A training task for deep learning models, typically transformers, during unsupervised pre-training. These tasks are not the final tasks the model will be evaluated on, but instead work to help the model gain an understanding of the organization and grammar of the genome (in the context of genomics). Commonly used pretext tasks are Masked Language Modelling (MLM) and Autoregressive Language Modelling (ALM)/Next Token Prediction (NTP)}
}
\newglossaryentry{Fully-connected layer}
{
        name=fully-connected layer,
        description={Also known as a densely connected layer. This is a neural network layer where each neuron is connected to every neuron in the previous and following layer. A fully-connected layer is often used as the final layer in a neural network to integrate learned features into predictions, leveraging its capacity to consider all feature combinations, which is critical for tasks like classification or regression}
}

\DeclareCiteCommand{\cite}[\mkbibsuperscript]
  {\usebibmacro{cite:init}%
  
  \iffieldundef{prenote}
     {}
     {\BibliographyWarning{Ignoring prenote argument}}%
   \iffieldundef{postnote}
     {}
     {\BibliographyWarning{Ignoring postnote argument}}}
  {\usebibmacro{citeindex}%
   \usebibmacro{cite:comp}}
  {}
  {\usebibmacro{cite:dump}}

\let\cite=\cite

\title{To Transformers and Beyond: Large Language Models for the Genome}

\author[1, 2, 3]{Micaela E. Consens}
\author[1]{Cameron Dufault}
\author[4]{Michael Wainberg}
\author[2, 5, 6]{Duncan Forster}
\author[2, 7, 8, 9]{Mehran Karimzadeh}
\author[7, 8, 9]{Hani Goodarzi}
\author[10, 11, 12, 13]{Fabian J. Theis}
\author[1, 14]{Alan Moses}
\author[1, 2, 3, 15*]{Bo Wang}
\affil[1]{Department of Computer Science, University of Toronto, Toronto, Ontario, Canada}
\affil[2]{Vector Institute for Artificial Intelligence, Toronto, Ontario, Canada}
\affil[3]{Peter Munk Cardiac Center, University Health Network, Toronto, Ontario, Canada}
\affil[4]{Prosserman Centre for Population Health Research, Lunenfeld-Tanenbaum Research Institute, Toronto, Ontario, Canada}
\affil[5]{Department of Molecular Genetics, University of Toronto, Toronto, Ontario, Canada} 
\affil[6] {The Donnelly Centre, University of Toronto, Toronto, Ontario, Canada}
\affil[7]{Department of Biochemistry \& Biophysics, University of California, San Francisco, San Francisco, California, USA} 
\affil[8]{Department of Urology, University of California, San Francisco, San Francisco, California, USA}
\affil[9]{Helen Diller Family Comprehensive Cancer Center, University of California, San Francisco, San Francisco, California, USA}
\affil[10]{Institute of Computational Biology, Department of Computational Health, Helmholtz Munich, Munich, Germany}
\affil[11]{TUM School of Life Sciences Weihenstephan, Technical University of Munich, Munich, Germany}
\affil[12]{Department of Mathematics, School of Computation, Information and Technology, Technical University of Munich, Garching, Germany}
\affil[13]{Munich Center for Machine Learning, Technical University of Munich, Garching, Germany}
\affil[14]{Department of Cell and System Biology, University of Toronto, Toronto, Ontario, Canada} 
\affil[15]{Department of Laboratory Medicine and Pathobiology, University of Toronto, Toronto, Ontario, Canada}

 \affil[*]{e-mail: bowang@vectorinstitute.ai}

\begin{abstract}
 In the rapidly evolving landscape of genomics, deep learning has emerged as a useful tool for tackling complex computational challenges. This review focuses on the transformative role of Large Language Models (LLMs), which are mostly based on the transformer architecture, in genomics. Building on the foundation of traditional convolutional neural networks and recurrent neural networks, we explore both the strengths and limitations of transformers and other LLMs for genomics. Additionally, we contemplate the future of genomic modeling beyond the transformer architecture based on current trends in research. The paper aims to serve as a guide for computational biologists and computer scientists interested in LLMs for genomic data. We hope the paper can also serve as an educational introduction and discussion for biologists to a fundamental shift in how we will be analyzing genomic data in the future.

\end{abstract}
\begin{document}

\flushbottom
\maketitle
\thispagestyle{empty}

\section*{Introduction}
In the past decade, deep learning has gone from a niche technology to a tool that has been successfully applied to a diverse range of tasks, from generating art \supercite{GLIDEAIart, DALL-E2}, language representation \supercite{BERT, GPT-2, XLNet, GPT-3}, and even to predicting protein structure from amino acid sequence \supercite{alphafold}. A common feature driving the success of deep learning tools at these tasks is the rapidly expanding size of available datasets, along with their improved accessibility and increasingly multi-modal nature. Perhaps just as important has been the push to generate larger and larger models \supercite{gpt4}. At the same time, novel techniques for capturing genomic information \supercite{wes, wgs} such as chromatin accessibility \supercite{atacseq, chipseq}, methylation \supercite{chipseq, methylseq}, transcriptional status \supercite{rnaseq, scRNAseq}, chromatin structure \supercite{hic} and bound molecules \supercite{chipseq} have provided a large and varied source of omics data to mine \supercite{encodeexplain}. 

The prevalence of deep learning tools applied to genomics in recent years is a natural consequence of more available and varied genomic data sources, as well as the introduction of larger deep learning models. Deep learning tools provide a solution to the computational and data analysis challenges posed by the increasing scale and variety of omics datasets\supercite{dLprimergenomics}. One of the key areas that deep learning has been applied to within the field of genomics is taking as input DNA sequences, RNA sequences, or single-cell RNA sequencing data and predicting missing high-dimensional modalities corresponding to the same data. For example, using DNA sequence, these models can predict regulatory annotations like transcription-factor binding, RNA-binding, chromatin accessibility, contact-maps, gene expression, RNA-seq coverage, promoter/enhancer region identification and more \supercite{DANN, deepBIND, deepSEA, deepFun, deeperbind, deepMILO, exPecto, deFine, ssl-dnn, xgboost, danQ, DNABERT, enformer, basenji, basset, diffgenedeep, splicingdeep, 3Dgenome, baseresmotif, JARVIS, DNABERT-2}. Using single-cell RNA sequencing data, these models can annotate cell types, correct batch effects, make gene dosage sensitivity predictions, among other tasks \supercite{scGPT, geneformer}.

Deep learning models have evolved over time to address the growing complexities of genomic data and the nuances of the many high-dimensional modalities available to measure the genome. For the last decade deep learning models on DNA sequence have been dominated by convolutional neural network (CNN) structures \supercite{deepBIND, deepSEA, deepFun, deeperbind, deepMILO, exPecto, deFine, danQ, basenji}. At the same time, deep learning models making predictions from single-cell RNA sequencing data have been dominated by different forms of autoencoders \supercite{single-cell-review, single-cell-review2}. However, as the popularity of transformer models \supercite{transformer} in the fields of computer vision and natural language processing (NLP), among others, has increased, so too has their application to genomic modelling problems \supercite{DNABERT, corigami, enformer, DNABERT-2, geneformer, nucleotide-transformer, scGPT, geneformer}. Transformer models have revolutionized NLP and various other fields, though they were initially conceived for sequence-to-sequence problems. Their applicability extends to genomic sequences, even being adaptable for tasks where the final goal of the model is not to produce a sequence but output a quantitative assay or classification \supercite{enformer, nucleotide-transformer}. Furthermore, transformers applied to genomic data offer a novel conceptual framework, the \textbf{attention mechanism}, to study the organization and grammar of the genome. Simultaneously, in the field of machine learning, computational advances continue to be made in improving the efficiency of the transformer, allowing the size of these models to increase along with their predictive power \supercite{efficienttransformer, improvementsinlongrangetransformer}. This will further accelerate the application of deep learning models for genomics, where methodologies for improving transformer efficiency are now being adopted in newer models, and most recently, genomic models are being proposed with novel architectures that claim to be the ``next transformer'' \supercite{hyena}. 

This review will discuss the trajectory of deep learning approaches in genomics, with a detailed discussion on the application, successes, and challenges of Large Language Models (LLMs) of the genome. LLMs, as referred to here, are models pre-trained on sequences of biological tokens such as DNA, k-mers, or gene identifiers. We adopt the term LLM, traditionally used in NLP, due to the pre-training approach analogous to that for text data. Although numerous review papers have explored deep learning models in genomics, with topics spanning from general introductions to specific discussions on model interpretation, understanding gene regulation, predicting the impact of genetic variation, and unveiling new applications \supercite{dLprimergenomics, DLcompbio2016, DLbiologyandmedicine2018, DLforbioinf2017, DLforneuroscience, DLforbiomedicine2018, explainableAI, interpretDLgenomics, deeplearningreview2023, deeplearningforgenomics, DLnewapplications, progressandchallenge}, we take a different approach. The paper will focus exclusively on the application to genomics of transformer-hybrid models, transformer-based LLMs, and other LLMs. Although non-transformer LLMs currently represent a nascent and less-explored niche, we believe that the exploration of architectures beyond traditional attention mechanisms will soon accelerate. The early research in this space holds promising and innovative prospects for the field. We will not discuss the use of transformers or language models for protein sequences or review non-LLM models used for single-cell transcriptomics as these topics have been reviewed elsewhere \supercite{protein-transformer-review, single-cell-review, single-cell-review2}.

In this paper, we first provide an overview of the transformer architecture, as well as a briefing on the Hyena layer, within the context of genomics. This is followed by a succinct review of previous deep learning architectures used in this field, establishing a foundation for an examination of novel architectures since. We introduce transformer-hybrid models, including transformers to predict assay data, followed by transformer-based LLMs and other LLMs applied to the genome. This review is intended for computational biologists with deep learning experience interested in understanding LLMs for the genome, as well as for computer scientists who are keen on gaining insights into the research opportunities within this exciting field.

\glsaddall
\printglossary[nonumberlist]

\begin{figure}
  \centering
  \begin{adjustbox}{max width=\textwidth}
  \includegraphics[width=\linewidth, totalheight=18cm]{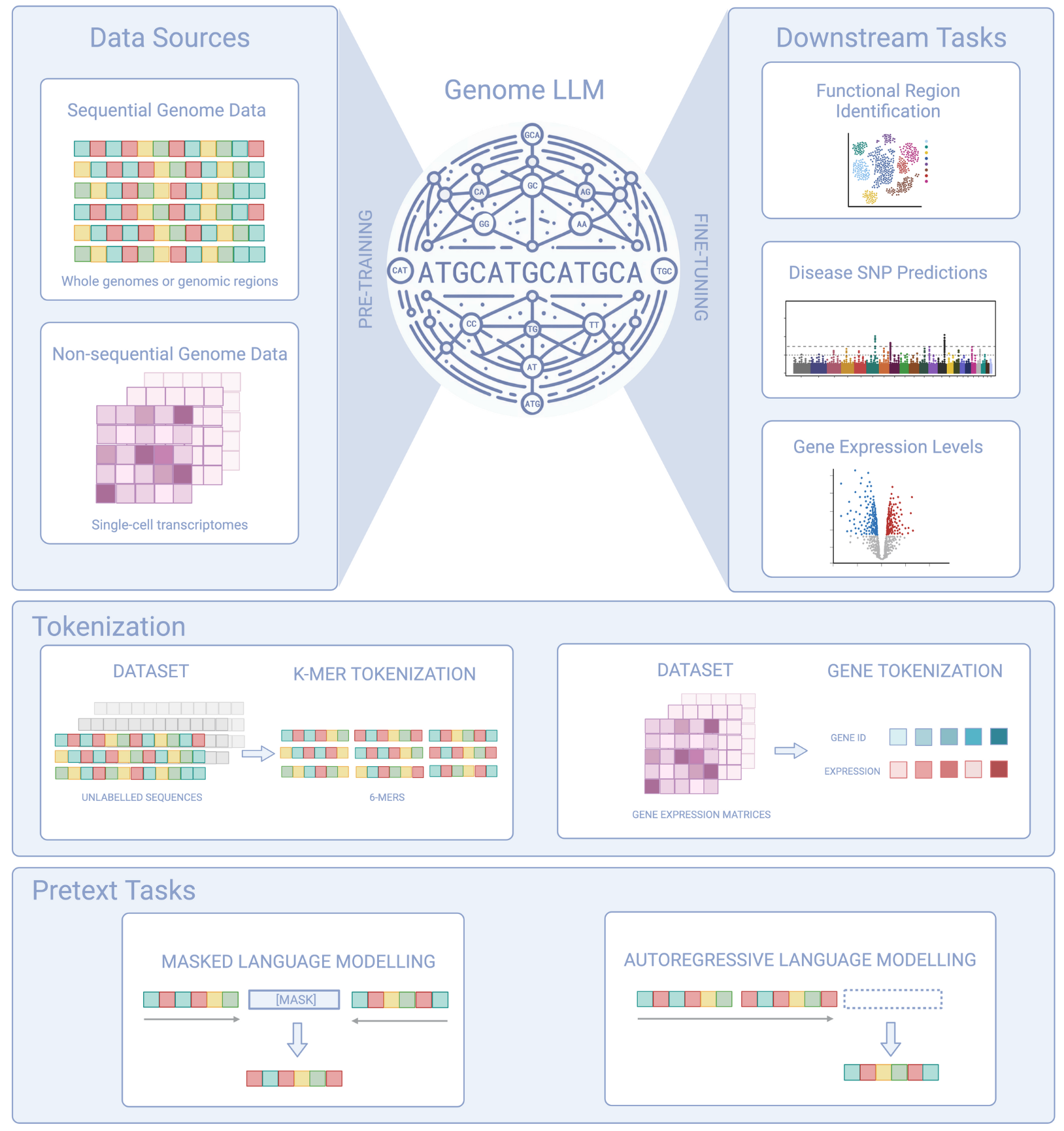}
  \end{adjustbox}
  \caption{\textbf{A big picture look at the power of Genome LLMs.} These models can take in both sequential (for example, DNA sequence but also ATAC-seq, Hi-C etc.) and non-sequential (for example, single-cell RNA-seq but also bulk transcriptome, multiome etc.) data and extract signals to make predictions on functional regions, disease-causing SNPs (single nucleotide polymorphisms or variations at a single position in a DNA sequence among individuals), gene expression predictions, and more. The training phase allows LLMs to learn the underlying structure of a dataset, and during the adaptation process (fine-tuning or prompting, etc.) the downstream task performance is evaluated. In the tokenization process different data sources (sequential vs. non-sequential) are prepared for input to the genome LLM differently. Sequence data is often tokenized using k-merization. In contrast, non-sequential data like single-cell RNA-seq data can undergo a more complicated tokenization scheme, where data can be represented using geneIDs as tokens, or some tokenization of gene expression values or rankings. LLMs often use the Masked Language Modelling pretext task in pre-training, where tokens are masked at different positions in the sequence. The model can use information from preceding and succeeding tokens to predict the masked token (as represented by the arrows indicating information flow available to the model). However, in the Autoregressive Language Modelling (ALM) pretext task the model sees a series of ``known'' tokens before being having to predict the next token. This means the model can only use preceding information (``known'' tokens) to make a prediction.}
  \label{fig1}
  \end{figure}
  
  \begin{figure}
  \centering
  \includegraphics[width=\linewidth]{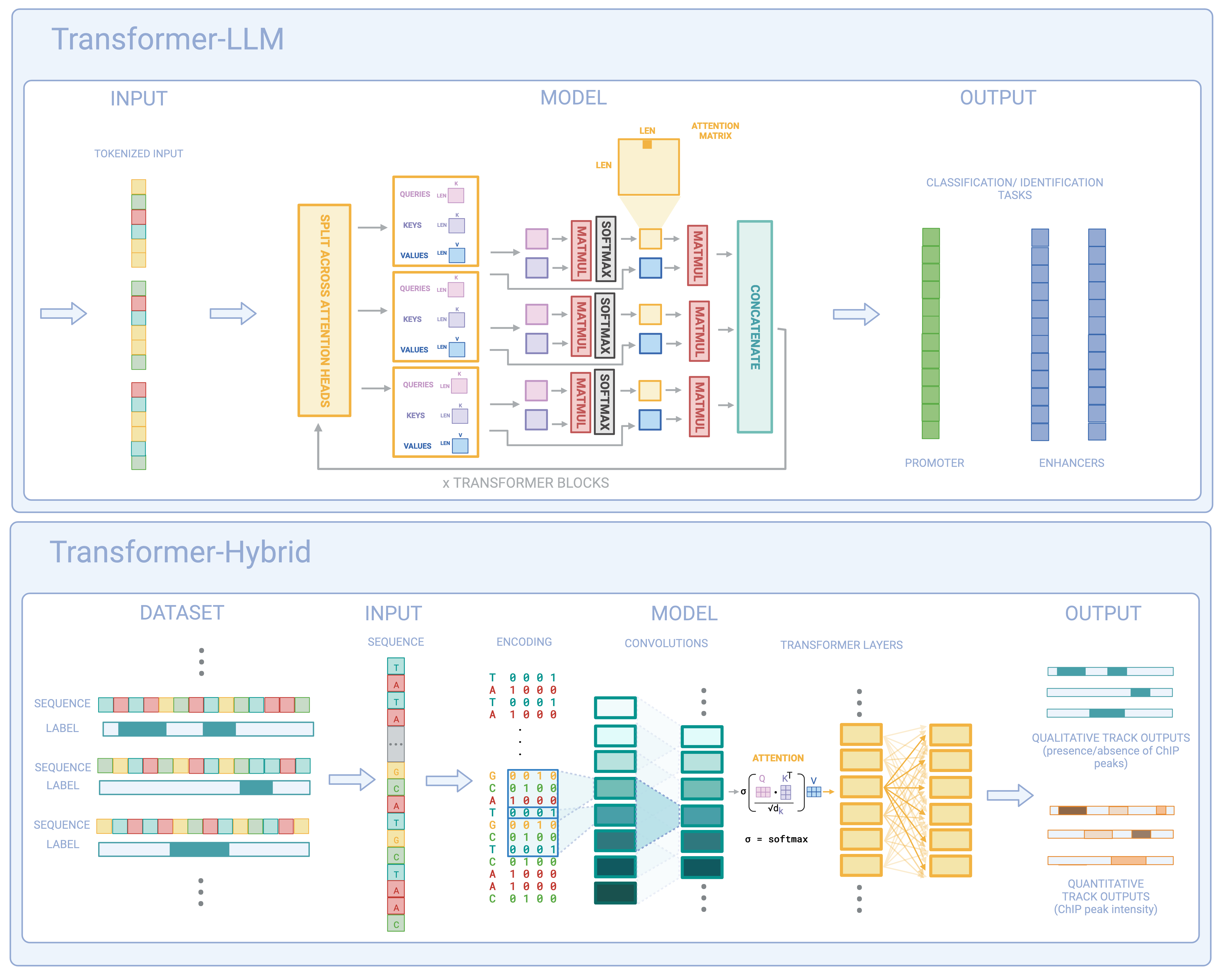}
  \caption{\textbf{Transformer-LLMs and Transformer-Hybrids. } Transformer-LLM models can take in tokenized data (either unlabelled sequence data or non-sequential data like single-cell RNA-seq data). The figure shows an example of k-merized DNA data, where k=6. Once tokenized, input data is immediately passed to a transformer module, the meta-layer combination of add-and-norm layers, fully-connected layers and the attention mechanism \supercite{DNABERT, nucleotide-transformer, scGPT}. In the figure, we focus on the attention mechanism within the transformer layer, outlining the steps involved in multi-headed attention. The keys, queries, and values in the attention mechanism are projections of the input sequence and attention is calculated across the inputted sequence at token-level resolution. Final predictions can be classification or identification tasks. The transformer-hybrid model takes in labelled data which is then one-hot encoded. Oftentimes, it is then passed through a series of convolutional layers, sometimes including dilated convolutions, before eventually reaching the transformer module. Here we represent the transformer module as a golden block, showing the attention mechanism is applied within these blocks. Transformer-hybrid models make predictions of experimental assays, either quantitative predictions like ChIP-seq intensities, or binary predictions like the presence/absence of ChIP-seq peaks \supercite{enformer, borzoi}.}
  \label{fig2}
  \end{figure}

\section*{Transformers}
To understand the use of transformers in the context of genomic modelling requires both a foundational understanding of the architecture and its usual training regime. In explaining the transformer, we assume readers have a grasp of several prominent concepts in machine learning, including the common layer types of deep feed-forward neural networks (DFNNs), convolutional neural networks (CNNs), and recurrent neural networks (RNNs). For a comprehensive introduction to deep learning, specifically in the context of genomics, we direct readers to the prominent review papers of Zou \textit{et al.}\supercite{dLprimergenomics}, Jones \textit{et al.}\supercite{DLcompbio2016}, Min \textit{et al.}\supercite{DLforbioinf2017}, and Wainberg \textit{et al.}\supercite{DLforbiomedicine2018}. If not already familiar, we encourage readers to familiarize themselves with DFNNs, CNNs \supercite{AlexNet}, and RNNs \supercite{RNN1, RNN2} using the existing literature. The glossary additionally provides concise definitions.

\subsection*{Architecture}
The transformer architecture, introduced in 2017, features layers of stacked \textbf{self-attention} mechanisms, typically \textbf{multi-head attention}, alongside addition and normalization layers, skip-connections, and fully-connected layers for final output predictions. We'll first explore the historical application of these components in deep learning for genomics before examining their integration within the transformer module as a ``meta-layer''. Following this, we'll outline the original presentation of the transformer as an encoder-decoder framework \supercite{transformer}, comprising stacked transformer-encoder modules and transformer-decoder modules. Finally, we'll review the various transformer model types applicable to genomics, including encoder-decoder transformers, encoder-only transformers, and decoder-only transformers.

\subsubsection*{Attention}
The first iterations of the attention mechanism were employed with Recurrent Neural Networks (RNNs), not in the context of transformers. However, computational efficiency was becoming a bottleneck for RNNs due to hardware constraints which prevented RNN-based models from expanding their parameter number \supercite{transformer}. The newly introduced attention mechanism of the transformer model eschewed recurrence, instead relying entirely on attention to draw global dependencies. The transformer's lack of recurrence was optimized for the attention mechanism to be computed in parallel on GPU hardware, overcoming the RNNs computational limitations. 

In the transformer, \textbf{self-attention}, or the use of attention to relate different positions within a single sequence to construct a representation of that sequence, is used to compute representations of input and output sequences. The transformers attention allows them to account for relationships within sequences that are not determined by immediate proximity, like RNNs, while leveraging parallelization on GPUs for computational efficiency, unlike RNNs \supercite{distalreg}. The key intuition in employing transformers for genomic sequences is their capacity to \textbf{attend} to long-range interactions within DNA sequences, which can span tens of thousands of base pairs. However, transformers still experience computational memory constraints. Direct base-to-base attention over very long-range genomic sequences (such as millions of base pairs) has not yet been achieved using transformers \supercite{enformer, DNABERT, corigami}. Instead, in genomics, attention is often applied to features extracted by dilated convolutions before transformer modules, thereby extending the range of self-attention calculations, at the cost of losing pairwise scores of relevance across a whole sequence \supercite{corigami, enformer, borzoi}.

Commonly used attention functions are calculated additively, or multiplicatively on sequences, where the latter case is referred to as scaled dot-product attention and is computed as follows\supercite{transformer}:
\[ attention(Q, K, V ) = softmax\left(\frac{QK^T}{\sqrt{d_k}}\right) V\]

Where the input sequence is transformed into the queries ($Q$), keys ($K$) in dimension $d_k$, and values ($V$) in dimension $d_v$. Here, queries represent the information at each position in the sequence, keys symbolize the associated information that the model should attend to, and values convey the information each position forwards to positions attending to it \supercite{enformer}. This setup facilitates the computation of the attention score as a weighted sum of all values, with the weight of each value being determined by the compatibility between the query and the corresponding key. The attention mechanism in transformers for genomic applications that use encoder-only transformers involves only ``self-attention''. ``Attention'' is calculated between the encoder-decoder, and as transformer models for genomics usually include only the encoder of the transformer model, there is usually only encoder self-attention. In these self-attention modules, queries ($Q$), keys ($K$), and values ($V$) all originate from the previous encoder layer, which ensures each position attends to all other positions in the prior layer.

Many transformer models employ a multi-headed attention mechanism, which facilitates information sharing across a sequence using fewer layers than vanilla self-attention \supercite{MHAexplain}. The attention calculation is modified as follows for $h$ attention heads with $d_{model}$-dimensional queries, keys and values \supercite{transformer}:
\[ multi\_head(Q, K, V ) = concat\left( head_1, ..., head_h\right) W^O\] where  \[head_i = attention(QW^{Q}_{i}, KW^{K}_{i}, VW^{V}_{i})\]
Where  $W^Q_i \in \mathbb{R} ^{d_{model} \times d_k} , W^K_i \in \mathbb{R}^{d_{model} \times d_k}, W^V_i \in \mathbb{R}^{d_{model} \times d_v}
$ and $W^O \in \mathbb{R}^{hd_v \times d_{model}}$. We can see the use of matrices of size $k \times n$ for each head shows that multi-head attention is a low-rank approximation for the full $n \times n$ attention matrix. 

Multi-head attention creates multiple different representations of the input to the transformer block through the learned representations of the inputs known as queries ($Q$), keys ($K$), and values ($V$). Instead of a single transformation of the input into queries, keys and values, the input is transformed into queries, keys and values for each of the $h$ attention heads in a separate calculation for each head. This means there are $h$ learned representations of queries, keys and values, or $h$ learned linear projections to $d_k$, $d_k$ and $d_v$ dimensions, respectively. The attention function is then calculated in parallel (very efficiently on GPU) for each of these learned representations to produce $d_v$ -dimensional output values. This approach lends more stability to the model during training and enhances modeling of long-range interactions \supercite{MHAexplain}. The value in multi-head attention is that it allows the model to jointly attend to information from different representation subspaces at different positions, as long as the different attention heads focus on distinct features, i.e. as long as the learned representations are not all the same \supercite{transformersurvey}.

\subsubsection*{Add-and-norm Layers, Skip-Connections and Fully-Connected Layers}
In transformers applied for genomics, add-and-norm layers perform a layer normalization and add in the original embeddings of a DNA sequence (for example) through a skip-connection. Normalization layers have long been used in deep learning as a way to prevent over-fitting, improve generalisation performance and accelerate convergence (a state during training in which the model loss becomes stable) \supercite{normalization}. Skip-connections were initially proposed as a solution to the Vanishing Gradient Problem, an artefact of backpropagation where the loss signal degrades proportionally to the depth of a network \supercite{resnet}. These layers increase the relationship between output signal and earlier layers in the model by providing connections between these layers.

Skip-connections have been widely used in genomics, such as in the Akita model \supercite{3Dgenome}. Akita was built by modifying the CNN-based Basenji architecture for the task of Hi-C contact domain prediction \supercite{hic, 3Dgenome}. The Akita model used dilated residual convolutions on DNA sequence data, introducing skip-connections, also known as residual connections, between successive dilated residual convolutional layers. Skip-connections used between dilated residual convolutions, as in Akita, allow for a model to learn to skip certain dilated convolutions across sequence if they do not contribute towards accurate prediction and instead hinder accuracy. 

Similarly, within transformers, skip-connections are used to solve the Vanishing Gradient Problem and, like in CNNs, have additional benefits. By encouraging the addition of the original embeddings through the skip-connection in the add-and-norm layer, information flow within the transformer mechanism is localized within the transformer block. While the self-attention mechanism continues to dynamically expand context based on relevance, the skip-connections remind the transformer what the original state was, ensuring the contextual representations of input tokens remain attached to those tokens.  

In transformers, fully-connected layers are used at the end of a transformer block to summarize information within a block. This information is either summarized for it to be sent to the next transformer block, or to be used for a final output.

\subsubsection*{Encoder-decoders}
Encoder-decoder frameworks are used outside of transformer architectures, and have found applications in genomics. The Orca \supercite{orca} paper leveraged an encoder-decoder framework composed of a one-dimensional (1-D) CNN encoder and a two-dimensional (2-D) CNN decoder to predict 3-D contact maps of the genome. The 1-D CNN encoder takes in a 1-D DNA sequence, and the 2-D CNN decoder outputs a 2-D contact map to make predictions of genomic proximity from sequence alone. The encoder-decoder framework proved particularly useful in the Orca \supercite{orca} paper, as using the encoder-decoder structure does not constrain input and output sequences to be the same length, accommodating the complexity and variability inherent in genomic data.

However, using a CNN-structure in an encoder-decoder framework, as used by Orca\supercite{orca}, limits the context window of a model to a predetermined kernel, or filter size. All encoded and decoded relationships are determined by the filter size applied to the sequence in the encoder and decoder. In transformers, however, the \textbf{attention} mechanism has potential to overcome this limitation. As long as the inputted sequence has full dense attention across it, all pair-wise relationships are calculated between sequence elements based on context.

The original transformer model was introduced as an encoder-decoder framework. The transformer-encoder is designed to encode input sequences into fixed-size context-aware representations, and the transformer-decoder produces variable-length sequences based on this representation and its own input. While the transformer modules in the encoder and decoder are very similar, the original transformer-decoder has an extra attention layer to focus on the encoder's output \supercite{transformer}. 

\subsubsection*{Encoder-only and Decoder-only Transformers}
Transformer models are not always encoder-decoders. Sometimes, transformer models can be used just as encoders, or just as decoders. In NLP, the Bidirectional Encoder Representations from Transformers (BERT) \supercite{BERT} and the Generative Pre-trained Transformer (GPT) \supercite{GPT-2} are two distinct frameworks for implementing transformers. BERT \supercite{BERT} is an encoder-only transformer framework with 12 transformer layers, each with 12 attention heads, usually pre-trained using a Masked Language Modelling (MLM) pretext task. MLM is a pretext training task where the model is pre-trained to predict a word that's been masked out of a sentence. The GPT framework \supercite{GPT-2} is decoder-only, and is pre-trained with an Autoregressive Language Modelling (ALM) pretext task. ALM is a pretext training task where the model is pre-trained to predict the next word in a sequence given the previous words. In NLP, GPT models generate text unidirectionally and sequentially, predicting the next word based on the previous one. BERT, in contrast, is designed for bidirectional representation, processing words from both left-to-right and right-to-left. In genomics, a GPT-style model would use preceding nucleotide information to make a prediction about subsequent nucleotides in a sequence. A BERT-style model could predict missing nucleotide information within a sequence using both preceding and succeeding nucleotides. 

BERT-based and GPT-based models work best for different tasks \supercite{BERT, GPT-3}. Encoder-only models, like BERT, are useful in the cases where final predictions have high accuracy when based off only an embedding, or feature representation of the inputted sequence. This is oftentimes true for classification tasks. BERT-based models pre-trained with MLM are useful for understanding genomic sequences where the overall context (both upstream and downstream) is important, such as identifying genomic features or classifying sequences where directionality is not critical. In genomic modelling, the BERT framework has already been applied. DNABERT \supercite{DNABERT}, as the model is called, can accurately identify whether DNA sequences are TATA or non-TATA promoters directly from the final embedding of the sequence alone. 

In NLP, decoder-only transformers pre-trained with ALM, like GPT-3 and GPT-4 \supercite{GPT-3, gpt4} are appropriate for tasks that involve predicting sequences where directionality is important. This includes modeling anything co-transcriptional or co-translational, such as RNA splicing or protein folding, where these sequences are biologically synthesized in a unidirectional manner. Decoder-only models in general provide the best \textbf{zero-shot generalisation} from purely unsupervised pre-training \supercite{decoder-onlyeval}. Zero-shot generalisation is a model's performance on tasks the model was not explicitly trained on, and unsupervised pre-training is a specific training regime where models are trained on large amounts of unlabelled data through a pretext task. Models exhibiting good zero-shot performance after pre-training on DNA sequences could capture multiple aspects of genomic grammar and structure instead of, for example, focusing on learning the best representation of a DNA sequence for predicting TATA-promoter regions (which may be too-constrained a task for good generalisation beyond this task). However, decoder-only models have yet to be widely adopted for transformers in genomics. A GPT-style model was implemented for single-cell omics data by scGPT \supercite{scGPT}, but it was not a true decoder-only model. The decoder-only framework for genomics has so far only been used by a model not based on the transformer architecture, HyenaDNA \supercite{hyena}.

\subsection*{Training}
The attention mechanism within transformers is often attributed as the main reason for their success \supercite{transformersurvey, transformersvision}. However, perhaps just as important as the attention mechanism, is the transformer's capacity to be pre-trained \supercite{ transformersvision}. While pre-training is not restricted to transformers, and has been shown to benefit other architectures immensely \supercite{pretrain-conv-vs-transformer, og-hyena}, transformers are the most commonly pre-trained architecture. 

In the field of genomics, as is the case in many areas of study, the availability of unlabelled data usually vastly exceeds that of labelled data \supercite{ssl-dnn}. The transformer model, along with some others, can address this imbalance by undertaking a pre-training phase, where it can learn in an \textbf{unsupervised} or \textbf{self-supervised} manner from a large unlabelled dataset through a pretext task. This pre-training stage effectively exploits the abundant unlabelled data available to learn general data representations. The model can then be fine-tuned for a specific task using a smaller labelled dataset, thereby benefiting from both the broad generalization capabilities gained during pre-training and the task-specific adaptations acquired during fine-tuning \supercite{transformersvision}.

% Interestingly, several transformer models have been applied to genomics without leveraging the benefits of this architecture's pre-training \supercite{enformer, borzoi}.

\subsubsection*{Pre-Training}
Pre-training can be unsupervised, supervised, or semi-supervised. Pre-training is an initial phase of training where a model, in this case a transformer, learns from a large corpus of data. When this data is unlabelled, and there is not a specific downstream task the model is being pre-trained for, the pre-training is called unsupervised. When the data has labels, the pre-training is considered supervised, and when both labelled and unlabelled data is used, the pre-training is considered semi-supervised.

Supervised pre-training can sometimes be preferred. This method can capture features that are more relevant to specific tasks a model will later be fine-tuned on, reducing the computational cost of fine-tuning. Supervised pre-training can sometimes lead to faster convergence or better performance on the downstream target task. Additionally, when there is a large amount of labelled data available for pre-training, it may not be necessary to use unsupervised pre-training.

Semi-supervised pre-training combines the benefits of both supervised and unsupervised learning, and their drawbacks as well. This approach can be beneficial when there is some amount of labelled data and complimentary unlabelled data available.

While each of these approaches to pre-training has their own strengths and limitations, unsupervised pre-training is an important direction for future research. This is especially true in biology, where much of the inherent signal in available data is not yet well enough understood to be labelled \supercite{futureunsupervised}. An unsupervised learning process allows a model to understand and capture general patterns and representations from the input data, independent of any specific downstream task. Additionally, this unsupervised approach potentially encourages the model to uncover signals in the data that researchers are unaware of. The broad pretext task assigned during the unsupervised pre-training phase, such as Masked Language Modelling (MLM) or Autoregressive Language Modeling (ALM)/Next Token Prediction (NTP), facilitates the model's learning of meaningful data representations despite not having labels. In the context of genomics, pre-training should expose the model to diverse genomic sequences, enabling it to understand various sequence patterns, context relations, and general nuances of genomic data. Pre-training results in a set of pre-trained weights that reflect these learned representations. These weights can then be updated during the subsequent fine-tuning phase as required.

\subsubsection*{Masked-Language Modelling}
MLM is a pretext task for the pre-training process. This technique involves randomly masking some of the input tokens, with the model's task being to predict these masked tokens. In the context of genomics, the `tokens' could be the individual bases in the DNA sequence. By masking out specific nucleotides within the genome and having the model ``fill in the gaps'', important relations between nucleotides are learned. The objective of this technique is to enable the model to understand the contexts in which each base can occur, thus gaining a holistic understanding of the genomic sequence. This exposure to a diverse set of sequences enhances the model's capacity to recognize different genomic sequence patterns and contexts. This is the most common pre-training task in genomics \supercite{DNABERT, GPN, DNABERT-2, nucleotide-transformer, geneformer}.

\subsubsection*{Autoregressive Language Modeling}
Autoregressive Language Modeling or Next Token Prediction is a training technique used in models such as Generative Pretrained Transformer or GPT \supercite{GPT-2,GPT-3}, where the task is to predict the next token in a sequence based on all previously observed tokens. In NLP, this usually means predicting the next word in a sentence. For genomic sequences, this could translate into predicting the next base in a DNA sequence given the preceding bases \supercite{hyena}.

This training regime allows the model to learn the probability distribution of a sequence and use this learned distribution to generate sequences. In the context of genomics, this could involve the model learning the probability distribution of bases in a DNA sequence, and then using this learned distribution to predict the likely continuation of a given DNA sequence. 

An artefact of this training is unidirectionality, meaning the model only considers previous, or ``known'', tokens in the sequence when predicting the next one. In contrast, bidirectional models based on BERT, like DNABERT \supercite{DNABERT} use MLM. They consider both previous and future tokens when predicting a masked token. As such, ALM and MLM provide different types of sequence understanding and can be useful for different types of tasks in genomics.

One key benefit of ALM in the context of genomics lies in sequence generation tasks, such as predicting the potential continuation of a partially known sequence. However, its unidirectional nature means it might be less effective at understanding the broader context of a sequence than bidirectional models based on BERT \supercite{BERT, DNABERT}, which consider both previous and subsequent tokens in the sequence. As such, the best choice of pre-training regime depends on the specifics of the task at hand. To the best of our knowledge, this type of pretext task has only been applied to the scGPT \supercite{scGPT} model in the context of transformers for genomics, and the HyenaDNA \supercite{hyena} model for the Hyena layer in genomics.

\subsubsection*{Fine-tuning}
Once the model is pre-trained, it undergoes a second training phase known as fine-tuning, a form of transfer learning. This process involves training the model on a smaller, task-specific dataset that is usually labelled. The pre-trained model parameters are adjusted (or ``fine-tuned'') during this phase, allowing the model to specialize in the target task. In fine-tuning, model weights from earlier layers learned in pre-training can be frozen, with only later layers adjusting their weights according to the loss function in the fine-tuning stage. Even if the previous weights are not frozen, it is expected that the model weights will not change very dramatically from the pre-training, only as much as is needed to accomplish the more specific task. Unfortunately, this is usually left as an assumption. The pre-training regimes of many genomic models are not evaluated closely to determine what the model is learning in pre-training vs. fine-tuning \supercite{BERTpretraining}. In theory, the fine-tuning process benefits from the broad representations learned during pre-training, enabling the model to effectively adapt to the specific task with less labelled data. For instance, in genomics, the unsupervised pretext task learned during pre-training could be MLM on the entire human genome, and the fine-tuning task could be classifying whether a particular sequence contains TATA promoters or not. The combination of pre-training and fine-tuning thus forms a powerful framework for learning from genomic data, balancing the benefits of unsupervised learning from large unlabelled datasets with supervised learning from smaller task-specific labelled datasets. 

\section*{Hyena}
Originally the Hyena layer type was introduced for NLP \supercite{og-hyena}, but more recently has been applied to genomic data in HyenaDNA \supercite{hyena}. The Hyena layer \supercite{og-hyena} was created to answer the demand for a transformer-like model to scale in terms of context length. The attention mechanism is bounded by quadratic cost in sequence length, and while adaptations have been made to improve the efficiency of the transformer \supercite{flashattention, efficienttransformer, improvementsinlongrangetransformer}, novel architectures have been proposed to replace it entirely. These new model architectures must be able to be pre-trained like the transformer and exhibit an attention-like mechanism, without the cost of attention. Model types claiming to be the ``next transformer'' include the linear transformer (an adaptation of the transformer to calculate attention in linear-time, likening it to an RNN) \supercite{lineartransformer}, RetNet (a linear transformer which removes the softmax in the attention calculation)\supercite{retnet}, and the Hyena layer \supercite{og-hyena}. 

The Hyena layer is so far the only one of these new model types to be applied to genomic data. The authors of Hyena \supercite{og-hyena} noted that there are two main properties of the attention mechanism that contribute to its success: global context, and data dependency. In attention, global context is achieved by having every token attend to every other token. The data dependency property of the attention mechanism is enforced by the softmax equation being applied to different projections of the inputted data. In contrast, a convolutional operator is not data dependent in this way, the same kernel is applied to every section of the inputted sequence. In attention, the input itself changes how the operation is applied.

The Hyena authors proposed that the global context and data-dependency properties of attention could be satisfied by another approach, one that would have better scalability in terms of context length. To achieve this, the authors introduced a subquadratic drop-in replacement for attention, composed of implicitly parameterized long convolutions and data-controlled gating \supercite{og-hyena}. An implicitly parameterized convolution is the opposite of an explicit convolution, in which there are $n$ parameters for the kernel size $n$. Instead, an implicit convolution is parameter-efficient. The parameters that define the convolution are learned by a Deep Feed-forward Neural Network (DFNN), such that there is a fixed parameter budget and any length kernel can be learned. A good parallel to this would be to learn the equation that defines a line ($y=mx+b$), instead of learning every point that must be plotted on the line. While this ensures the Hyena layer is parameter-efficient, it does not explain how the layer is sub-quadratic in cost. Long convolutions cannot be naively applied, even implicit ones, as a convolution with a kernel size equal to the input length would result in a quadratic-cost computation like attention. To remedy this, the Hyena authors use the Fast Fourier Transform \supercite{FFT} to bring the cost of computing a long convolution down, essentially moving the convolution calculation from the time domain (with timesteps $t$) to the frequency domain, or the Fourier domain.

\begin{figure}
  \centering
  \includegraphics[width=\linewidth]{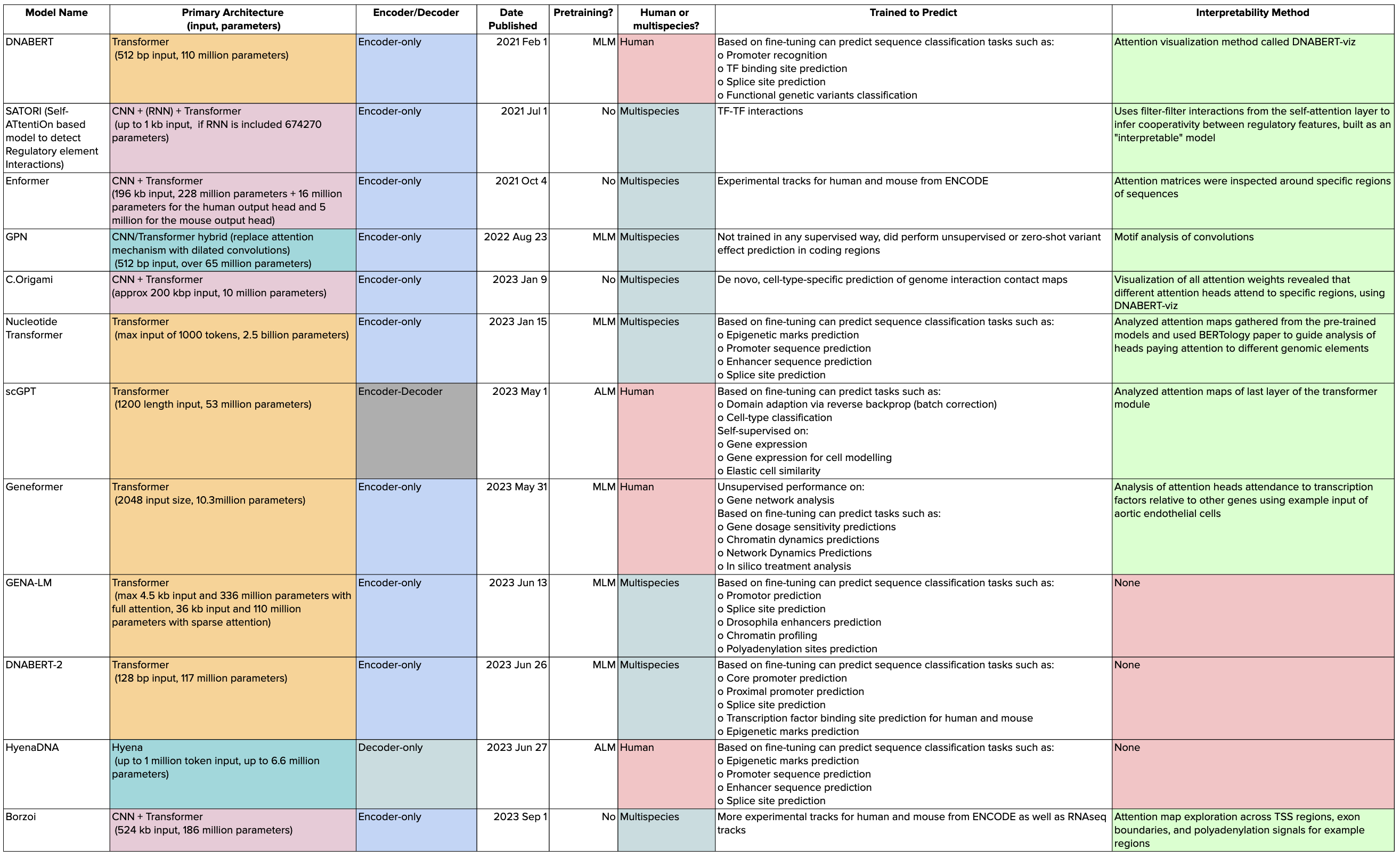}
  \captionof{table}{\textbf{A summary of the recently proposed deep learning models covered in this review.} In the architecture column the yellow colour indicates a transformer LLM model, the pink colour indicates a transformer-hybrid model, and the turquoise colour indicates a non-transformer LLM. The encoder-decoder column in colour-coded by encoder-only models, decoder-only models and encoder-decoder style models. The datasets trained on column has a red colouring for human-only models and blue for cross-species training. The interpretability column is coloured by presence or absence of model-interpretability analysis.}
  \label{tab1}
  \end{figure}

\section*{Deep Learning for Genomics}
For a more general review on deep learning models applied to genomic data we suggest readers look into Li \textit{et al.} \supercite{deeplearningreview2023} and Eraslan \textit{et al.} \supercite{deeplearningforgenomics}. Readers interested in the applications as well as progress and challenges of these tools should read Routhier \textit{et al.}\supercite{DLnewapplications} and Spoval \textit{et al.}\supercite{progressandchallenge}. For more background on model interpretation in genomics we direct readers to Novakovsky \textit{et al.}\supercite{explainableAI} and Talukder \textit{et al.}\supercite{interpretDLgenomics}. The main focus of this section will be to explore the transformer model and similar architectures (including HyenaDNA), and will discuss the previous architectures only as context for these model's application in genomics.

\subsection*{Before the Transformer}

\subsubsection*{CNNs}
CNNs offer a powerful tool for sequence analysis and the identification of complex regulatory patterns \supercite{CNNsurvey}. Using convolutional layers, CNNs exploit spatial hierarchies and the concept of weight sharing, enabling efficient detection of regulatory patterns and motifs regardless of their genomic location (within a certain kernel size) - a feature known as translational equivariance. The genomic sequence information is captured in convolutional layers by a series of operations with learned kernels that match specific DNA subsequences, demonstrating a versatile and adaptable method for processing genomic data\supercite{AlexNet, understandingCNN, conceptCNN, explainableAI}. 

Early CNN models for genomics, like DeepBind \supercite{deepBIND}, were simple architectures by today’s standards. Subsequent models, such as DeepSEA \supercite{deepSEA} and Basset \supercite{basset}, increased model depth through additional convolutional layers. These models further expanded their context size, allowing them to process longer sequences and specialize in complex tasks like chromatin accessibility prediction. Later, the introduction of dilated convolutions, such as in the Basenji \supercite{basenji} model, marked a significant leap in the receptive field size of genomic CNNs. Dilated convolutions greatly expand a CNN model’s context size, enabling the capture of long-range dependencies without increasing a model’s computational complexity \supercite{basenji, basenji2, enformer}. 

Many of the CNN-based models for genomics predict specific experimental assays given a DNA sequence. However, the Akita \supercite{3Dgenome} model offers a unique perspective on DNA-sequence based prediction. Akita aims to model the three-dimensional (3D) structure of the genome by leveraging a CNN structure with dilated convolutions. Akita transforms one-dimensional (1D) DNA sequence data into two-dimensional (2D) representations indicative of genomic region contacts. The major innovation of the Akita model was the use of the dilated residual convolutions. Residual connections allow for increased context window size by optionally adding motifs captured from earlier layers back into later ones. 

The Orca model \supercite{orca}, which succeeds the Akita in 3D genome prediction, implements an encoder-decoder framework using CNNs. The success of this approach motivates the use of classic encoder-decoder type frameworks in genomics.

In single-cell transcriptomics modelling, the CNN has not been used widely, though it has been applied for de-noising data, by converting single-cell data into images first \supercite{sc-CNN-GAN, single-cell-review2}. 

More recent CNN-based models for genomics have introduced modules beyond vanilla, dilated and dilated residual convolutions. The ExPecto \supercite{exPecto} model includes a spatial transformation module to reduce the dimensionality of genomic predictions. Dimensionality reduction is a way to simplify data by reducing its number of features or ``dimensions'', which can be done by removing redundancy or useless information. In ExPecto this is done through weighing different regions of the sequence captured by the CNN module based on their relative distance to a transcription start site (TSS). Sequences closer to the TSS are weighed as more important. This dimensionality reduction helps the ExPecto model by reducing the model's likelihood to overfit (memorize the data) and lessening the computation time and storage requirements for the data. Additionally, the spatial transformation module allows for incorporation of spatial information from a larger context window, continuing with the general theme of genomic modelling \supercite{dimRed}. 

The ExPecto model's spatial transformation module provides a strong indication of the potential success of transformers applied to the same problem. The use of attention mechanisms in transformers allows for dynamically selecting subsequences as important from a larger region, rather than statically assigning importance based on sequence proximity to the TSS. Similarly, the Hyena \supercite{og-hyena} layer's ability to take long convolutions over whole sequences with data-controlled gating expands sequence context while reducing the dimensionality of the sequence information.

\subsubsection*{RNNs}
When RNNs were introduced to genomic sequence prediction, they were, for the most part, employed in combination with CNN modules \supercite{danQ}. As most RNNs have a ``memory'' feature, they are aptly suited for processing sequential data and capturing dependencies within a sequence \supercite{dLprimergenomics} (see Supplemental Figure \ref{suppfig1} (c)). 

An initial limitation of RNNs was the degradation of this ``memory'' over long sequences, which spurred the development of Long Short Term Memory networks (LSTMs). These networks introduced ``smart'' neurons that selectively ``remember'' or ``forget'' information, thereby overcoming the limitation of memory degradation inherent to some standard RNNs. Bi-directional LSTMs (BLSTMs) build on this technology, integrating dual-direction processing to incorporate information from both preceding and succeeding sequence elements. This is particularly relevant in genomics due to the lack of inherent directionality of the genome \supercite{dLprimergenomics} (see Supplemental Figure \ref{suppfig2} (b)).

While the RNN enjoyed a lot of success in the NLP field, it was not broadly adopted for genomic modelling \supercite{danQ, deepMILO, sc-RNN-cellmot, single-cell-review2}. Perhaps this was due to the imminent introduction of the transformer model, which used attention to replace recurrence, allowing for parallelized computations on GPU hardware that made the model easier to scale. Interestingly, several recent adaptations on the transformer block aiming to increase context-size have re-introduced recurrence to their architecture, as the attention mechanism is becoming a bottleneck for compute and memory \supercite{retnet, og-hyena, lineartransformer}.

\subsection*{The Transformer}

For genomic modeling, transformers are employed in two primary ways: either as a subsequent module following initial layers, or as a standalone transformer model.

In the former, a set of initial layers precede the transformer modules, these layers compress broad contextual data into a lower-dimensional embedding space. This approach is often necessitated by the quadratic computational cost of attention mechanisms. 

When an extended context window is not a requirement for downstream tasks, transformers can operate as standalone models. Inputs are transformed using a limited number of distinct tokens and processed directly by the transformer module. Models that adopt this approach are primarily aimed at creating Large Language Models (LLMs) tailored for genomic sequences.

Therefore, we split the transformer-based models for genomics into two classes: hybrids and LLMs. Hybrid models, which incorporate transformers as one element within a more intricate architecture, may not meet the criteria for being considered true LLMs. Unlike LLMs, which are designed for comprehensive understanding and generation of language-like sequences, hybrid models are specialized for tasks such as predicting experimental assays like Cap Analysis Gene Expression (CAGE) tracks and ChIP-seq. These models pick up where the traditional genomic models, such as DeepBind, left off. They specialize in higher accuracy performance on many of the same tasks as their CNN predecessors \supercite{deepBIND, deepSEA}. Hybrid models often forgo pre-training, a hallmark feature commonly associated with LLMs. This lack of pre-training further distinguishes them from LLMs, which are generally more versatile and generalizable. 

\subsubsection*{Hybrid Transformers: Assay Prediction}
SATORI (Self-ATtentiOn based model to detect Regulatory element Interactions) is a transformer-based model. The model aims to capture a global view of the interactions between regulatory elements in a sequence to infer cooperativity. This model uses a convolutional layer in addition to a self-attention mechanism (see Figure \ref{fig2}). The convolutional layer is used to discover motifs in the input sequences, while the self-attention layer is used to capture potential interactions between these features. This approach allows the model to infer a global landscape of interactions in a given genomic dataset directly, rather than as a post-processing step.

In SATORI, the sparsity inherent to the attention matrix allows it to be used as a proxy for covariance. Additionally, the model was proposed as ``interpretable'' due to direct analysis of the attention matrix. However, treating the attention mechanism within transformers as directly interpretable has drawbacks \supercite{og-lrp}. 

Within a few months of SATORI, the Enformer model was published. Enformer uses transformers to predict various genomic track signals including gene expression, DNA accessibility, and histone modification or TF-binding information \supercite{enformer}. The convolutional blocks and transformer blocks within Enformer work together to dynamically summarize and capture long-range dependencies in DNA sequence. The model's final layers bifurcate into organism-specific network heads for prediction, having been trained on both human and mouse data.

Enformer represents an evolution in the prediction of quantitative genomic assays. The model moves away from the traditionally utilized convolution-based architectures such as Basenji \supercite{basenji} and ExPecto \supercite{exPecto}. Although convolutional models have shown efficacy in capturing DNA sequence patterns, they, like the transformer, can also suffer from a limited context window. Naive long-convolutions are as computationally intensive as attention when the filter size of the convolution is equivalent to the sequence length. While dilated convolutions can mitigate this by introducing ‘holes’ in the filter, there is still a limit to the expansion of the CNN context size. Enformer’s innovation is balancing the use of convolutions, dilated convolutions, and transformers to maximize context length while also capturing much of the local sequence information.

Recent work by Karollus \textit{et al.} \supercite{currsequencemodels} scrutinized models like Enformer and Basenji2 \supercite{basenji2}, emphasizing the need for more and better-curated training data. They noted that despite Enformer's dramatically increased context window, it still encountered considerable limitations in predicting the impact of distal regulatory elements, such as enhancers. Enformer's predictive power remains comparably robust even with a severely restricted input window, suggesting its receptive field size is not the sole determinant of its success. This success can instead be attributed to innovations in model architecture such as combinations of various layer types, overall parameter number, or the quantity of data it was trained on. Karollus \textit{et al.} \supercite{currsequencemodels} propose that in order for models to accurately account for distal regulators' contributions they must train on datasets curated with an emphasis on long-range signals.

Borzoi \supercite{borzoi} builds on the Enformer architecture by doubling the size of the context window. Borzoi further expands the number of experimental assay predictions compared to Enformer, and introduces predictions of RNA-seq coverage. The model’s main innovation is using an architecture styled after U-Net to upsample and increase resolution prediction following the transformer module. The convolutional blocks preceding the transformer summarize the longer sequence input and transform it into the same resolution as in Enformer. This allows the attention calculation to be calculated with similar compute, thereby avoiding the quadratic memory complexity by increasing sequence resolution. To make final predictions at a higher resolution, the information captured by the transformer is expanded to a larger sequence length using convolutions.
 
C.Origami is the hybrid transformer iteration on 3D genome prediction \supercite{3Dgenome, orca, corigami}. The model makes de novo predictions of cell type-specific chromatin architecture from sequence. C.Origami incorporates both DNA sequence and genomic signals (CTCF-binding and ATAC–seq) to do this. Like its predecessor, Orca \supercite{orca}, C.Origami uses an encoder-decoder design, with the modification of using two encoders to deal with multi-modal input types (one for DNA sequence and one for genomic signals). The novelty of this design also includes following the dual-encoders with a transformer module before the task-specific decoder. After the 1D-CNN encoders, genomic feature representations from both encoders are concatenated and fed into the transformer module to enable long-range information exchange. This, along with the addition of genomic features, allows C.Origami to outperform Orca \supercite{orca, corigami}, along with Akita \supercite{3Dgenome}. The C.Origami model enables in silico experiments that examine the impact of genetic perturbations on chromatin interactions, and moreover, leads to the identification of a compendium of putative cell type-specific regulators of 3D chromatin architecture. 

\subsubsection*{Transformers: Large Language Models}
One of the earliest Large Language Models (LLM) for the genome also happens to be one of the earliest uses of a transformer applied to the genome. DNABERT was adapted for genomic sequence modelling \supercite{DNABERT} based on the original BERT model \supercite{BERT}. DNABERT is trained using a method similar to BERT \supercite{BERT}, with some modifications to suit the genomic context. The model is pre-trained on overlapping k-mers of genomic sequence using unsupervised learning with the MLM pretext task. This pre-training phase allows the model to learn a rich representation of the genomic sequence data, capturing both local and long-range dependencies.

After pre-training, DNABERT is fine-tuned on specific tasks, such as predicting the presence of transcription factor binding sites or other regulatory elements. This involves training the model on a smaller amount of labeled data, with the objective of predicting the labels based on the input sequence. The fine-tuning phase allows the model to adapt the representations learned during pre-training to the specific task at hand. The DNABERT model, as an LLM, does not directly predict regulatory annotation measurements. Instead, DNABERT predicts the absence or presence of binding sites, promoters, etc. given a specific classification objective it is fine-tuned on.

DNABERT has state-of-the-art performance in predicting proximal and core promoter regions (two specific kinds of promoter types), and presence of transcription factor binding sites. However, the model has a more limited context window relative even to previous CNN and RNN-based models. Like the original BERT, the model cannot take in more than 512 tokens, or in this case, k-mers. To mediate this limitation, the authors designed DNABERT-XL, a version of DNABERT where the authors split longer sequences into multiple pieces, each under 512 base pairs in length, and independently feed these pieces to DNABERT. They then concatenate the final representations together and feed the concatenated representations to the output layer. While this approach to increasing context-window size was able to identify between TATA and non-TATA promoters well, they did not demonstrate an end-to-end approach in modeling complex long-range dependencies, due to the limitations of the cost of attention.

The major advantage of DNABERT lies in its capacity for self-supervised pre-training, a product of its transformer architecture. As seen in Table \ref{tab1}, the model is pre-trained extensively before any fine-tuning tasks. This alleviates the need for large amounts of labelled task-specific data, or extensive training time and compute spent when fine-tuning the model for specific downstream tasks.

DNABERT's k-merization scheme for tokenizing the genome was adopted by many transformer-based LLMs for the genome, including the Nucleotide Transformer \supercite{nucleotide-transformer}. The Nucleotide Transformer consists of a family of transformer models with varying training regimes and parameter sizes. These models include a 500 million parameter model trained on the human genome, a 500 million and a 2.5 billion parameter model trained on a large set of genetically diverse human genomes, and a 2.5 billion parameter model trained on genomes from hundreds of species. Each of the Nucleotide Transformer models were trained on hexamer versions of their genomic datasets and pre-trained using the MLM pretext task, like DNABERT. However, the Nucleotide Transformer models replaced the overlapping k-merization of DNABERT with a non-overlapping version that greatly reduced tokenized sequence length. The replacement of overlapping k-mers with non-overlapping ones also addressed a major limitation of overlapping k-mers in MLM tasks, where information from a masked token is leaked by adjacent tokens \supercite{DNABERT-2}. However, the non-overlapping k-mer approach comes with its own limitations, primarily that the insertion or deletion of a single nucleotide base leads to dramatic changes in tokenized sequences \supercite{DNABERT-2}. This dramatic change in representation of sequences, despite their similarity, prevents the model from easily aligning proximal sequences within the token space, which can impede the ease of training.

The smallest of the Nucleotide Transformer models is five times larger than DNABERT, and the authors' benchmarking experiments (predicting enhancers, promoters, TATA promoters, splice sites etc.) show that increasing model size yields better performance. This is in line with the same intuition that has led assay-prediction models like Enformer and its predecessors to continue increasing their parameter sizes. The results of the Nucleotide Transformer models’ benchmarking experiments also showed that training with intra-species variability (using multiple genomes of a single species, such as thousands of human genomes) did not perform as well as training with inter-species variability (their multi-species training regime). This could be due to the multi-species models better capturing functional importance conserved across evolution, allowing them to generalize better even on human-based prediction tasks. The Nucleotide Transformer models strongly suggest that models that leverage evolutionarily diverse data in pre-training will have improved capacity in capturing functional relevance.

The next iteration of DNABERT, DNABERT-2 \supercite{DNABERT-2} (released by the same group as DNABERT), followed a multi-species training regime. DNABERT-2 also utilized an alternative tokenization scheme, replacing k-mer tokenization entirely, instead favouring a compression algorithm used more widely by LLMs in NLP known as Byte Pair Encoding (BPE) \supercite{BPE}. This approach bypasses the issues associated with overlapping tokenization (leakage), and non-overlapping tokenization (unnecessary distance in the token space between similar sequences). The BPE approach iteratively merges frequent pairs of nucleotides or segments within the genome, as opposed to using a specific k-mer. This results in the model's vocabulary comprising a set of variable-length tokens that together represent the entire genome dataset across species. BPE tokenization of DNA sequences has been observed to result in biologically significant tokens, with the longest tokens corresponding to elements of the genome known to be repetitive \supercite{genalm}. This is in contrast with the arbitrary tokens utilized when k-mer tokenization is used. Furthermore, the BPE method remains as computationally efficient as non-overlapping tokenization.

Along with the introduction of BPE instead of overlapping k-merization, the DNABERT-2 authors employ several other methods to improve computational efficiency over DNABERT, including the use of Flash Attention \supercite{flashattention}, among others \supercite{ALiBi, hu2021lora}. Flash Attention \supercite{flashattention} is an IO efficient manipulation of the attention mechanism, optimizing for read and write time in memory as well as improving computational efficiency by splitting the keys, queries and values into blocks that have softmax incrementally performed over the entire input. These modifications allow DNABERT-2 to perform comparably to the Nucleotide Transformer models in several tasks, despite 21 times less parameters and significantly less computational cost.

Another recently introduced family of transformer-based DNA LLMs is GENA-LM \supercite{genalm}. Like DNABERT-2 it uses BPE tokenization, and like Nucleotide Transformer there are both human-only and multi-species models with varying amounts of parameters. However, a significant difference between GENA-LM and other models is the use of sparse attention to help mitigate the quadratic complexity in context length of the transformer's attention mechanism. Sparse attention allows for modelling distant dependencies within the sequence, but does not perform full pairwise attention between all tokens. This results in GENA-LM models having increased maximum sequence length over other transformer-based LLMs for the genome, with a maximum tokenized sequence length of 4096 tokens. The median token length after BPE tokenization is nine base pairs, thus GENA-LM models can process sequences of up to 36 thousand base pairs.

\subsubsection*{Transformers: Non-sequential Genome LLMs}
Within the family of transformer-based LLMs for the genome, there are two recently proposed models that have taken a radically different approach to modelling the genome. Specifically, these models depart from modelling DNA or RNA in a sequence-based manner, instead training on non-sequential single-cell omic data such as cell profiles and gene expression counts. 

The first of these models is Geneformer\supercite{geneformer}, which operates on non-sequential single-cell omic data. Geneformer represents cell embeddings with only gene expression ``rankings'', ignoring precise gene-expression measurements and instead treating single-cell omic data as qualitative. Geneformer was trained on a 30 million cell corpus, consisting of only healthy human cells. The input to Geneformer  consists of a rank value encoding, such that genes are inputted in order of their expression ranking within the cell (normalized across the entire single-cell corpus). Geneformer's input has a length limited to 2,048 tokens, still making it the largest input with full dense self-attention for transformer-only models in genomics so far. These 2,048 input tokens are the rank encodings of all genes within a given cell that have non-zero expression. The attention matrix is used to learn gene-to-gene relationships within the context of each cell, over the full 2,048 gene input. 

This approach to reporting the presence or absence of genes within a cell, and ranking the present genes within the embedding, benefits from observing gene expression values for a single cell many times across the 30 million cell corpus. The authors state this approach encourages highly expressed housekeeping genes to be normalized to lower ranks, while genes that identify cell state with high fidelity, but which may be expressed at low levels, move to higher rankings. The ranking approach may even have benefits over reporting gene expression values quantitatively by offering a buffer against technical artefacts that bias these counts.  

The Geneformer model reports excellent performance in gene dosage sensitivity predictions, chromatin dynamics predictions, and network dynamics predictions.

Like Geneformer, the recently proposed single-cell Generative Pretrained Transformer model, or scGPT \supercite{scGPT},  takes as input non-sequential genome data. However, unlike Geneformer, scGPT uses a unique tokenization scheme to handle single-cell transcriptomic data and marks one of the first uses of generative pre-training for transformers applied to genomics. The input to scGPT is described analogously to word and sentence embeddings used in generative NLP where cells (sentences) are characterized by genes and the protein products they encode (words). Each input embedding of length 1,200 to scGPT therefore represents a cell composed of gene IDs, their binned expression values (which converts all expression counts into relative values), and condition tokens representing additional learned features or meta-information. Including metadata tokens allows scGPT to accept as input multiple single-cell omic data modalities, or single-cell multi-omic data, including scATAC-seq data. Prior to scGPT's approach there was no concrete way to aggregate multiple single-cell data modalities in one foundation model such that biological signal was preserved and experimental noise was reduced. 

Along with scGPT's novel tokenization scheme, the author's introduced a novel attention mask for the generative pre-training stage. scGPT's generative pre-training was based on the GPT \supercite{GPT-3, gpt4} generative pre-training, where models aim to predict a next most likely token based on a ``prompt'' of known inputs. The motivation for using a generative approach with ``prompting'' to model single-cell omic data, beyond zero-shot generalization advantages, was twofold. scGPT was designed to solve both the problem of generating unknown gene expression values given known gene expression values (``gene prompts''), as well as generating whole genome expression given a known input cell type condition (``cell prompts''). In order to achieve this, scGPT required a unique attention mask capable of supporting gene-prompt and cell-prompt generations. This was achieved by only supporting attention computations between embeddings of ``known'' genes and the query gene itself. As the generation iterations continue, the predicted gene expression values for the new set of query genes become the ``known'' genes for the next iteration. This forces sequential predictions in non-sequential single-cell omic data, with parallels to causal attention masking in vision. 

In contrast, scGPT's fine-tuning training regime employed a non-generative approach, instead using MLM for optimal performance. scGPT, pre-trained on 33 million healthy human cells, achieves state-of-the-art performance in genetic perturbation prediction, batch correction, and multi-omic integration, further demonstrating the power of ``pre-training universally, fine-tuning on demand''.

\subsubsection*{Beyond the Transformer}
While the transformer model is often considered synonymous with Large Language Models, especially with the recent success of ChatGPT \supercite{GPT-3, gpt4}, LLMs are not always transformer-based. There are other architectures that can be trained to perform as well as LLMs, and can even undergo similar pre-training regimes including leveraging MLM or ALM pretexts \supercite{pretrain-conv-vs-transformer, og-hyena}. As of now, it remains unclear whether the success of the transformer model lies in an artefact of the architecture, like the attention mechanism, or whether this mechanism simply allowed these models to scale up more quickly than their counterparts. It could be that the pre-training capabilities of the transformer, which are not restricted to this architecture, contribute immensely to its success. If this is the case, the transformer has the potential to be overtaken in NLP, and genomics as well \supercite{og-hyena, lineartransformer, retnet}.

The Genomic Pre-trained Network, or GPN \supercite{GPN}, copied the exact architecture of a transformer encoder module, but replaced the attention mechanism with a convolution operation across the sequence. The idea behind this came from recent work that showed pre-trained CNNs are competitive with transformers in NLP \supercite{pretrain-conv-vs-transformer}, and protein modelling \supercite{convproteins}. The GPN model leverages the MLM pretext task in pre-training, and is trained solely on individual nucleotides, rather than using BPE or any k-merization strategy. The genomes used in pre-training consisted of eight Brassicales reference genome assemblies from NCBI Genome. Instead of sampling the whole genome equally in 512 base pairs windows during pre-training, the authors took the union of exons (with a small intronic flank), promoters (1,000 base pairs upstream of the TSS) and a complimentary number of random windows from the whole genome. While the authors state this may have improved performance, they do not show any experiments to validate this claim.

The authors demonstrate GPN learns non-coding variant effects from unsupervised pre-training on genomic DNA sequence alone, outperforming supervised deep learning models such as DeepSEA.

Another non-transformer genome LLM, HyenaDNA\supercite{hyena}, achieves a context size of 1 million nucleotides, 500x larger than the largest of the foundational models utilizing full pairwise attention, the Nucleotide Transformer\supercite{nucleotide-transformer}. Instead of relying on the quadratic-bound attention mechanism, which compares each pair of points in a sequence, the authors of the original Hyena paper designed a subquadratic-time layer. This layer is constructed by interleaving implicitly parameterized long convolutions and data-controlled gating. The Hyena layer incorporates several artefacts from CNNs, RNNs and transformers. The long convolution is essentially a convolutional layer performed with a filter/kernel size equal to the sequence length. The long convolution is implicitly parameterized, meaning the filter is represented as a parametric function of the time step. This allows the model to capture dependencies across the entire input sequence without a prohibitive increase in computational cost or risk of overfitting. The data-controlled gating works similarly to the ability of LSTMs to ``forget'' certain information, learning a function that computes the gating values based on the input data. From the transformer, the Hyena layers have borrowed the idea of global interaction mapping, and taken the famous pre-training regime associated with transformers, but swapped the quadratic complexity of attention for long convolutions and data-controlled gating. HyenaDNA is based off of the structure of the decoder-only transformer architecture, replacing the attention mechanism directly with the Hyena operator. HyenaDNA is trained generatively using the ALM pretext task, like scGPT \supercite{scGPT}. The HyenaDNA model was only trained on one reference human genome, providing an obvious direction for future work. All in all, the model boasts state-of-the-art performance on all eight datasets from GenomicBenchmarks \supercite{genomicbenchmarks}.

% \section*{Timeline of Computational Approaches}
% \input{timeline.tex}

% \section*{Interpretability}
% \input{interpretability}

\begin{figure}
  \centering
  \includegraphics[width=\linewidth]{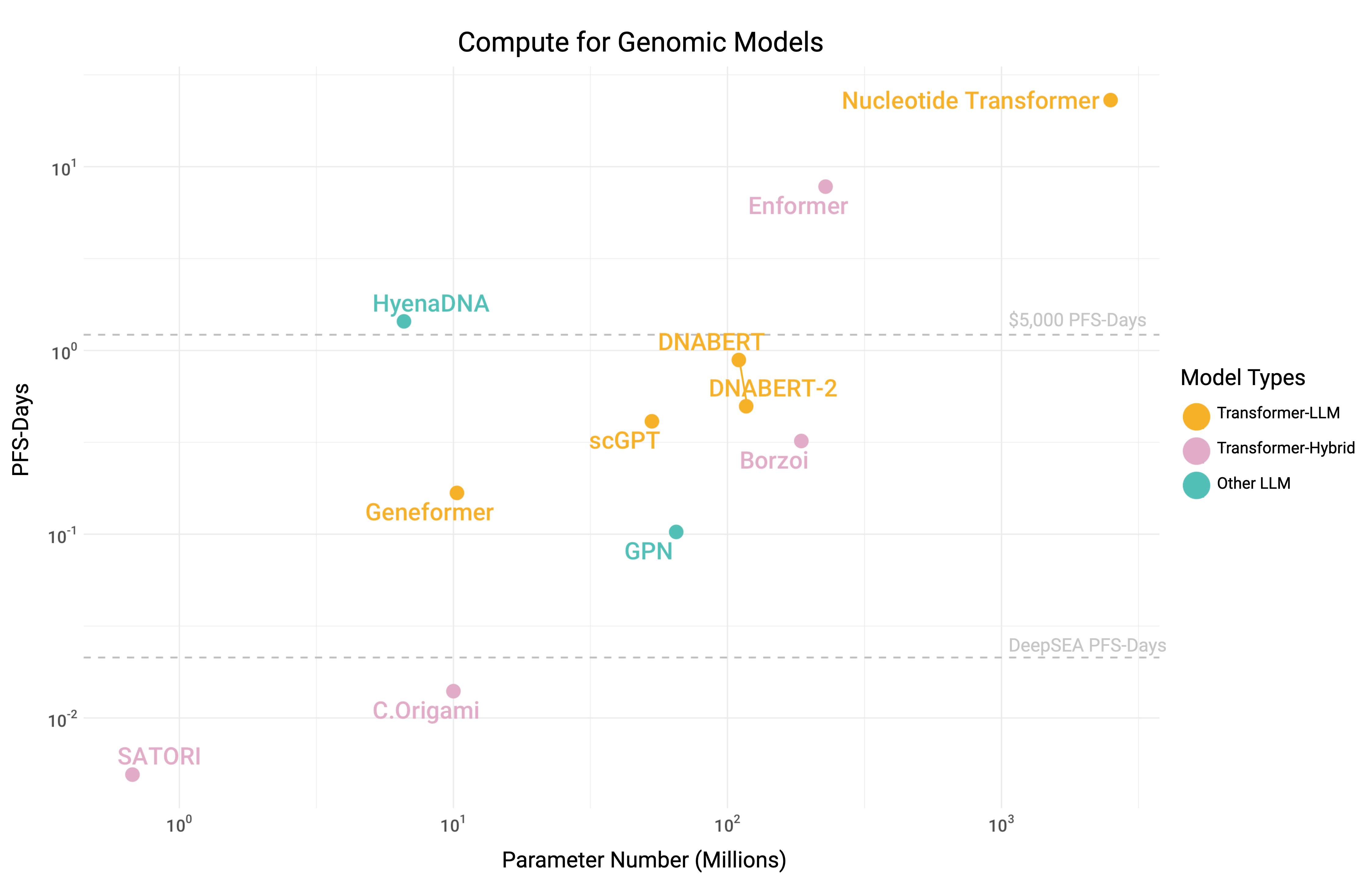}
  \caption{\textbf{The total amount of compute, in petaflop/s-days (PFS-Days) used to train the various models discussed in the review (all of the models for which parameter number, training time, and GPU usage was available).} A PFS-Day consists of performing 1015 neural net operations per second for one day, or a total of about 1020 operations. It is a compute-time measurement proposed by OpenAI to compare across model architectures which can be thought of similarly to kW-hr for energy. For context, we calculate the PFS-Days for the original DeepSEA model as well as the equivalent PFS-Days that can be purchased to train a model with \$5,000 USD, renting 8 A100 GPUs at \$8.80/hour. Calculations for PFS-Days can be found in the Limitations section.}
  \label{fig3}
  \end{figure}

\section*{Limitations}
% Transformer-based models have emerged as powerful tools for predicting regulatory annotations from genomic sequences. These models leverage the transformer architecture, which has revolutionized the field of NLP and machine vision, and now is poised to do the same for genomics. However, despite this new architecture's potential, the transformer model itself, and recent transformer-based models, present several limitations and challenges that must be addressed.

% Most importantly, the tasks and the manner in which transformers are being trained for these tasks, will not solve regulatory annotation prediction. Solving the problem of regulatory annotation prediction does not only mean having perfect scores in predicting ENCODE or Roadmap data. A model that has truly captured the regulatory grammar of the genome can unveil long-range regulatory dependencies within the genome, and work across various cell-types. While impressive, and indicative of a promising future in regulatory annotation prediction, the results of current models like DNABERT \supercite{DNABERT}, SATORI \supercite{satori}, Enformer \supercite{enformer}, and C.Origami \supercite{corigami}, also showcase how far we still are from solving the problem of regulatory annotation prediction. 

% other papers blah blah blah about the limitations of cell-type specificity etc, we are focus on the specific transformer nad beyond limitations

As previous review papers have focused on the limitations inherent in applying deep learning models to genomic data, including cell-type specificity debates and training data limitations \supercite{deeplearningforgenomics}, we focus on the limitations inherent to the application of novel architectures to genomics, such as the transformer and Hyena.

\subsubsection*{Long Range Interactions}
Transformer-hybrid models, despite high accuracy on specific tasks for which large amounts of labelled training data exist, still appear unable to capture long-range dependencies within the genome. Most improvements in assay prediction models are aimed at increasing context window in an effort to better model these long-range dependencies. However, Karollus \textit{et al.} \supercite{currsequencemodels} suggest a larger receptive-field might not be the key factor driving success in more recent models, including Enformer\supercite{enformer}. While trends show models with increased context window size have had higher accuracy on predicting experimental assays, this does not necessarily suggest context window size is the driving force behind improved predictive capacity. Karollus \textit{et al.} showed that significantly reducing the input size given to the Enformer model has a minimal effect on its performance, suggesting that transformer-based models, like Enformer, may achieve state-of-the-art performance simply due to the addition of the transformer module, or through increased parameter size. Enformer's successor, Borzoi \supercite{borzoi} appears to better integrate long-distance information, as measured by ranking distal regulatory elements for their gene-specific enhancer activity (data from CRISPR screens). However, there have not yet been experiments done to evaluate Borzoi's \supercite{borzoi} performance after a significant reduction of the input size, similar to those performed on Enformer \supercite{enformer}. Likely, context window size has to be combined with better-curated datasets, ones that are curated to capture the effects of distal eQTLs, distal enhancers and distal repressors.

If such a dataset can be curated, then context-window size will be the driving factor for modelling long-range dependencies. In that case, transformer-based LLMs, without being preceded by downsampling techniques to reduce dimensionality prior to the attention calculation, or applying a more efficiently implemented attention method \supercite{flashattention}, will be limited by their context window. Even the Nucleotide Transformer \supercite{nucleotide-transformer}, with 2.5 billion parameters, was only able to extend the context window to a max of 1,000 tokens, which still remains 1,000 times smaller than the HyenaDNA \supercite{hyena} context window. 

Other-LLM models that forgo the use of the quadratic attention mechanism have potential to better capture long-range interactions in the genome.  As long as the attention mechanism itself is not the driving force of success in these models, which does not appear to be the case given the recent success of HyenaDNA \supercite{hyena}, this is a strong avenue for potential research.

\subsubsection*{Cell type Specificity}
Transformer-hybrid models, which aim to predict experimental assays, are usually trained with either ENCODE \supercite{ENCODEDHS} data from multiple cell lines, or fine-tuned non-specifically across cell types. The tasks they are evaluated on make predictions ignoring the inter-cell type variability that exists within genomic annotations. While this allows the models to leverage huge amounts of data available on ENCODE \supercite{ENCODEDHS} and Roadmap \supercite{ROADMAP} by pooling cell types together, many findings in the field of genomics and the increased used of single-cell specific sequencing suggests there is cell-specific heterogeneity in regulatory annotations like chromatin accesssibility, chromatin conformation, gene expression, and transcription factor binding \supercite{cellhetereogeneity, singlecellatac, deFine}. Some transformer models have recently been proposed to mitigate the bias of cross-cell lines gene expression prediction using transfer learning, but this approach has yet to be commonly adopted \supercite{transferchrome}. A future area of research could be moving away from transformer-hybrid models for predicting cell-type specific experimental assays. Instead, the prompting capabilities of generative LLMs could be leveraged to create cell-type specific contexts for predictions.

\subsubsection*{Data Privacy}
As mentioned, the majority of the models discussed in this review are trained on public datasets like ENCODE \supercite{ENCODEDHS} and Roadmap \supercite{ROADMAP}. However, there is a risk of re-identification of individuals even from anonymized genomic datasets due to the uniqueness of genetic information. As more data continues to be integrated into these models, and these models are potentially leveraged for use on private datasets including in clinical settings, there is an increasing need for developing and implementing more robust de-identification techniques and privacy-preserving algorithms, such as differential privacy \supercite{differentialprivacy} and federated learning \supercite{federated-learning}.

\subsubsection*{Interpretability}
Beyond limitations in capturing long-range dependencies, a fundamental limitation of applying deep learning models to genomic data lies in their black-box nature. As these models become larger and more complex, it becomes less obvious how to explain and understand the decisions they make, especially in the context of genomics where the underlying ``language'' of the genome is as of yet unclear to us \supercite{interpretDLgenomics}.  Previous papers have explored the interpretability of deep learning models in the context of genomics extensively \supercite{deeplearningforgenomics, explainableAI, dLprimergenomics}, here we focus specifically on the task of interpreting transformer models and similar architectures. 

Attention scores of transformers have been proposed as an interpretability solution to the problem of black box models in genomics \supercite{transformersexplaingenomics, enformer, DNABERT}. This had led to models like Enformer \supercite{enformer}, DNABERT \supercite{DNABERT} and C.Origami \supercite{corigami}, and more, directly reporting analysis of attention scores from their models as proof of their ability to capture biological signal. However, several studies on transformer models outside the context of genomics have shown attention is not inherently interpretable \supercite{uninterpretableattention, transformerinterpret}.  Specifically, reporting raw attention scores alone reduces the information captured by the model to only the attention scores, ignoring most of the attention components by only accounting for the inner products of queries and keys, rather than the entire computation of queries, keys and values. Furthermore, the attention scores taken from a specific layer in the model ignore other layers involved in the model's decision-making process. Aggregation across multiple layers in a naive summation or mean does not account for the way attention scores move through the model, through add-and-norm layers and skip-connections \supercite{transformerinterpret, transformer}. Models that consider the mean of attention heads across multi-headed attention also dilute the information captured by the model as different heads contribute differently in each layer, and not all heads contribute equally \supercite{uninterpretableattention, transformerinterpret, multiheadlrp}.

Recently, other methods like attention flow and attention rollout \supercite{attentionflowandrollout}, along with Layer-Wise Relevance Propagation (LRP)\supercite{transformerinterpret, multiheadlrp} have been applied to interpret transformer models with success.

The attention rollout method makes the assumption that input tokens are linearly combined through the layers of a transformer according to their attention weights. The weights are then rolled out to capture the propagation of information from input tokens to hidden embeddings in a given layer $i$, recursively multiplying the attention weights matrices. Attention flow treats the problem of mapping the contribution of input token attention weights from final layers as a maximum flow problem. In other words, the attention weights are treated as the capacity of each connection between neurons, where computing the attention in a layer $i$ back to the inputs is treated as finding the maximum flow value from each input token to each position in layer $i$. Both methods, attention rollout and attention flow, allow for quantification of the flow of information through self-attention in a transformer. This allows the attention weights within the model to be correctly mapped back to the input tokens and given a measure of their relative relevance to model outputs. These methods are highly correlated with one another and have been shown to outperform raw attention scores alone in determining token contributions to outputs\supercite{attentionflowandrollout}. However, the rollout approach is limited by treating input contributions as neutral rather than distinguishing between positive and negative contributions of tokens to final decisions. Additionally, the attention flow method is computationally expensive.

LRP is a method which has been utilized to understand and interpret the decisions made by deep learning models like CNNs \supercite{og-lrp}. LRP works by attributing the contribution of each input feature to the final decision of the network. This is achieved by propagating the output prediction back to the input layer, thereby providing an indication of feature importance.

LRP is calculated as \supercite{towardexplainableai}:
\begin{equation}
\label{eqn:LRP}
R_j =\mathlarger{\mathlarger{\sum}}_{k}\frac{a_j w_{jk}}{\mathlarger{\sum}_{0,j} a_j w_{jk}}R_k
\end{equation}

Where $j$ and $k$ are the neurons of consecutive layers, $a$ is the activation of the respective neuron, and $w$ is the weight between two neurons. $R_k$ is then the ‘relevance’ received by neuron $k$, which is interpreted as the contribution of that neuron in its layer to the output prediction $f (x)$.

LRP has been widely used in interpreting CNN-based models, providing valuable insights into their decision-making process \supercite{og-lrp}. LRP has also been applied to transformers, particularly to demonstrate multi-headed attention results in ``redundant'' heads \supercite{multiheadlrp}. A recent paper has further adapted the LRP method for explaining transformer classification decisions, incorporating information from attention scores across multiple heads and re-weighting contributions by relevance\supercite{transformerinterpret}. This method has outperformed other attribution methods like classic LRP \supercite{og-lrp2}, partial LRP \supercite{multiheadlrp}, rollout \supercite{attentionflowandrollout} and Grad-cam \supercite{gradcam} on transformers.

Beyond adapting interpretability methods for transformers, model-agnostic methods like SHAP (SHapley Additive exPlanations) \supercite{shap-val}, and more recently, weightedSHAP \supercite{weighted-shap-val}, provide obvious avenues of exploration. SHAP quantifies the contribution of each feature to a model's prediction on a specific data point. WeightedSHAP extends this by considering the importance of each data point, useful when some instances matter more than others. 

Ultimately, interpretation of transformers in genomics beyond attention-score visualization has been limited. We acknowledge that while previous papers provide some insight into transformers' interpretability specifically in genomics \supercite{transformersexplaingenomics}, they have not compared methods beyond attention scores or acknowledged the limitations of this method \supercite{uninterpretableattention}. 

Outside of the transformer models covered in this review, the other LLM models must be evaluated for their interpretability in the context of genomic tasks. GPN  \supercite{GPN} reported motifs captured in the convolutional blocks of the architecture as a metric of interpretability, similar to previous CNN-based models for genomics \supercite{deepSEA, deepBIND, deeperbind}. These motifs were then compared to experimentally determined and validated motifs and reported as Position Weight Matrices (PWMs) or logos. Interestingly, despite HyenaDNA's \supercite{hyena} success in genomic modelling, the authors did not dedicate a section of their paper to model interpretability. This is likely due to the novelty of the Hyena layer, as interpretability methods have not yet been explored for this architecture. Potential routes to address this issue could involve examining the motifs captured by the model and used for the ``gating'' mechanism within the Hyena layer's ``memory''. 

The transformer-based models covered in this review have reported attention scores for their interpretability metrics, with the SATORI \supercite{satori} authors even claiming the model was inherently interpretable due to the attention mechanism alone. While these models have not employed more sophisticated methods for model interpretability, attention mechanisms appear to encourage exploration of model interpretability. In contrast, only one of the non-transformer LLM models (GPN \supercite{GPN}) reported some kind of model interpretability. We suggest researchers in this area who want to employ the use of transformer-based models analyze their models beyond raw attention scores alone, by incorporating methods like classic LRP \supercite{og-lrp2}, partial LRP \supercite{multiheadlrp}, SHAP \supercite{shap-val} or weightedSHAP values \supercite{weighted-shap-val} and rollout \supercite{attentionflowandrollout} to discover biological motifs and tokens of interest these models attend to. This will potentially result in richer and more nuanced explanations of genomic models that use attention mechanisms, and could lead to more meaningful biological insight. Furthermore, we encourage the concurrent development of interpretability methods for the novel architectures proposed to this field beyond transformers, like HyenaDNA \supercite{hyena}.

We expect the use of interpretability methods for attention-based mechanisms in genomics to increase, given their increasing usage trend among transformers applied to other fields \supercite{transformerinterpret}. Similarly, we expect models that aim to unseat the transformer in genomics invest in interpretability methods along with novel architecture design.

\subsubsection*{Compute Requirements} 
Perhaps the most noteworthy limitation in the usage of LLMs for the genome, irrespective of the use of transformer, convolution, or Hyena layer, is the computational cost of training these models. Their training regimes, pre-training in particular, often necessitate high-performance computing resources, which may not be readily accessible or affordable for all research teams. 

Using a compute calculation derived by OpenAI, we can calculate the peta-flop(s)/days for each of the models discussed in the review, with the exception of GENA-LM \supercite{genalm} for which there was insufficient data on days trained, in order to compare across different model architectures and parameter numbers.  The equation for PFS-Days using GPU Time is: 

\[ \text{PFS-Days}  = \text{Number of GPUs} \times (\text{peta-flops}/\text{Hardware}) \times \text{Days Trained} \times \text{Estimated Utilization}\]

Where OpenAI assumes a 33\% utilization for GPUs. Looking at Figure \ref{fig3} we can see that the majority of models proposed using transformers or similar architectures for genomics can be trained with \$5,000 USD on 8 A100 GPUs. However, of the largest and arguably most discussed models (DNABERT \supercite{DNABERT}, Enformer \supercite{enformer}, Nucleotide Transformer \supercite{nucleotide-transformer} and HyenaDNA \supercite{hyena}) only DNABERT \supercite{DNABERT} can be trained with \$5,000 USD worth of PFS-Days. However, DNABERT was trained prior to 2021, when GPU access was more restricted and more expensive than it is today, meaning the model likely could not have been trained using the equivalent of \$5,000 PFS-Days at that time.

The transformer and transformer-based models reviewed in this paper necessitate significant processing power and memory due to their complex architecture, which includes multiple layers of multi-headed attention mechanisms. Even the Hyena model, which boasts much better time and memory complexity than the attention mechanism, requires more compute than most academic labs can afford (setting \$5,000 USD as a baseline). Additionally, even with compute available, transformer-based models and HyenaDNA as well, can take a considerable amount of time to train, see Figure \ref{fig2}. This can slow down the research process and make it more difficult to iterate and improve models.

For pre-training tasks, even when leveraging unsupervised learning, these large models typically require large amounts of training data to perform optimally. In the context of genomics, this could mean needing access to extensive datasets of genomic sequences, of which there is not currently a lot of cell type specific data. While unsupervised pre-training can mitigate the need for labelled data, many of these models require fine-tuning to perform well on downstream tasks, and due to the size of the models, still require significant amounts of labelled data to fine-tune effectively.

\subsubsection*{Pre-training Task Design} 
Pre-training is a powerful and architecture-agnostic tool for genome LLMs, but its effectiveness hinges on the quality of the task assigned for pre-training. At best, pre-training allows deep learning models to capture universal data patterns \supercite{oppsandrisksoffoundationmodels}, at worst, it is an unnecessary expenditure of compute \supercite{limitszeroshot}. To maximize the benefits of pre-training for a genomic LLM, the initial task should be both biologically relevant and useful for later applications.

Consider a genome LLM, $G$, fine-tuned to predict disease outcomes from DNA sequences, $G(X)=Y$. Using common pre-training tasks like ALM may impose flawed assumptions, such as treating DNA sequences as unidirectional. While MLM accounts for DNA's bi-directional nature, it's not without its own limitations.

Often, pre-training tasks from other fields like natural language processing (NLP) are applied to genomics \supercite{selfgenomenet} without consideration of their biological relevance and the inherit differences between the genome and natural language. These tasks may not capture meaningful biological signals or align with the model's ultimate goal. A better approach is to design a pre-training task based on biological insights. For example, pre-training $G$ to estimate the evolutionary viability of a mutated sequence would be more aligned and scientifically sound. One recently proposed technique for biologically informed unsupervised pre-training of a deep learning model on genomic data is Self-GenomeNet\supercite{selfgenomenet}. Self-GenomeNet exploits the structure of genomic data by utilizing the reverse-complement of sequences during training, with the pretext task being to predict the embedding of the reverse-complement of the sequence which follows a given sequence. Thus, the model learns to embed in a representation of a given sequence the information necessary to reconstruct the reverse-complement of the neighbouring sequence. Self-GenomeNet has been demonstrated to perform well for data-scare genomic tasks.

Ultimately, the genome LLMs discussed in this review have applied ALM or MLM pretext tasks for pre-training, with minimal adaptations for biological context. The effectiveness of these models pre-training has been largely unexplored, though initial investigations on the performance of pre-trained models or on pre-training regimes for genome LLMs have not been favourable \supercite{limitszeroshot,BERTpretraining}. While this remains a limitation of current genome LLMs, it also provides a promising direction for future research.

\section*{Future Directions}
The success of deep learning models for the genome, specifically with the increasing use of LLMs, and the limitations they are currently bounded by, provide a complex outlook for the future. Of the notable trends within deep learning genomic modelling, one of the most prominent is the potential of unsupervised pre-training regimes, and specifically of multi-species pre-training. This approach could capture evolutionarily conserved data in the genome and better model its underlying grammar. Furthermore, this approach transcends the need for specific architectures like the transformer, or the Hyena layer. However, with the success of many of these models taken as contingent upon their expensive and time-consuming pre-training regimes, it is important to understand what exactly these models are learning in pre-training vs. fine-tuning. Recent work \supercite{BERTpretraining} investigating BERT model behaviour in genomics shows that k-mer embeddings from random data have comparable performance on downstream tasks to k-mer embeddings pre-trained on real biological sequences. Further work examining the zero-shot performance of models like scGPT and Geneformer suggests designing more biologically meaningful pretext tasks could greatly remedy their limitations \supercite{limitszeroshot}. This tells us that while pre-training and unsupervised learning have potential to increase the power of genomic models, the pre-training tasks for these models must be well designed and validated to prove true genomic grammar is being captured. We suggest further experiments are conducted on these models to compare pre-trained and fine-tuned vs. randomly initialized and fine-tuned embedding spaces. Additionally, while unsupervised pre-training provides an excellent direction for future research, we wonder if pre-training on the entire genome is the best way to leverage the power of pre-training. Repetitive non-coding sections of DNA make up nearly half of the human genome \supercite{TEs}, and could potentially overpower the ability of these models to learn more relevant signals from relatively less common but more important sequence regions. 

New LLM architectures like the Hyena layer are emerging, which don't rely on attention mechanisms yet still support pre-training. These models may offer better scalability for genomic data compared to traditional transformers\supercite{retnet, GPN, hyena}. The attention mechanism's quadratic complexity is a bottleneck for genomic modeling, especially when increasing the context window size is crucial. This opens the door for next-gen models like HyenaDNA to potentially outperform transformers\supercite{hyena}.

While these newer models have their own drawbacks, mainly in interpretability, they pose a real challenge to transformers. If the transformer's key advantage has been its scalable context window size, it will need significant improvements to stay competitive. Efforts are underway to refine attention mechanisms, such as introducing sliding windows\supercite{neighbourhoodattention}, enhancing efficiency\supercite{efficienttransformer, ToME}, and improving long-range interaction modeling in NLP\supercite{improvementsinlongrangetransformer}. Despite these advancements, transformers still lag behind in context window size when compared to models like the Hyena layer.

GPT-4's prevalence and power continues to raise more interest in creating LLMs for the genome, however, it is even more interesting to consider the field will move away from the LLM entirely. Large models that can integrate multi-modal data could unify genomic, transcriptomic, proteomic, and epigenomic data, offering a more holistic view of biological systems than LLMs can provide on their own. If trends in deep learning models for proteins are predictive of future trends in genomic sequence modelling, it is very likely the next model to try for the genome will be the Diffusion model \supercite{diffusion}. If pre-training with evolutionarily varied data is a key to success in modelling genomic information, perhaps diffusion, which naturally models evolution, is the obvious choice. A recent paper by Alamdari et. al. \supercite{protein-diffusion} marries these concepts together for protein generation, showing evolutionary sequence data might be all we need, to great success.

The future of powerful and interpretable deep learning models in genomics is one of personalized medicine, understanding evolutionary dynamics, drug discovery, synthetic biology, and more. We are at an extremely exciting time for the field, and we hope to see an increase in the use of multi-species pre-training, and more biologically motivated and downstream-task aligned approaches to designing pretext tasks. We believe research in this area will lead to the greatest success in modelling the genome, whether it be through transformer models, Hyena layers, or diffusion models. Additionally, it is our belief that deep learning models will only succeed in modelling genomic data if strides continue to be made toward their interpretation within genomic contexts \supercite{transformerinterpret, bertology}.

% \section*{Limitations and Future Directions}
% \input{limitationsfuture}

\printbibliography

\bigskip
\noindent\textbf{Acknowledgements}\\
We acknowledge the support of the Natural Sciences and Engineering Research Council of Canada (NSERC). All figures were created with BioRender.com.

\noindent\textbf{Author contributions}\\
M.E.C. selected the papers to review, summarized contributions from all papers, performed analysis and designed all figures. A.M., B.W., M.W., and D.T.F. helped with figure design. C.D. contributed to paper selection and summarizing contributions and A.M., M.W., M.K., F.J.T., and H.G. contributed to manuscript writing. A.M. supervised and B.W. conceived and supervised the project.

\noindent\textbf{Competing interests}\\
The authors declare no competing interests.
\clearpage

\section*{Supplemental Figures}

\begin{figure}[!ht]
  \centering
  \includegraphics[width=\linewidth, totalheight=19cm]{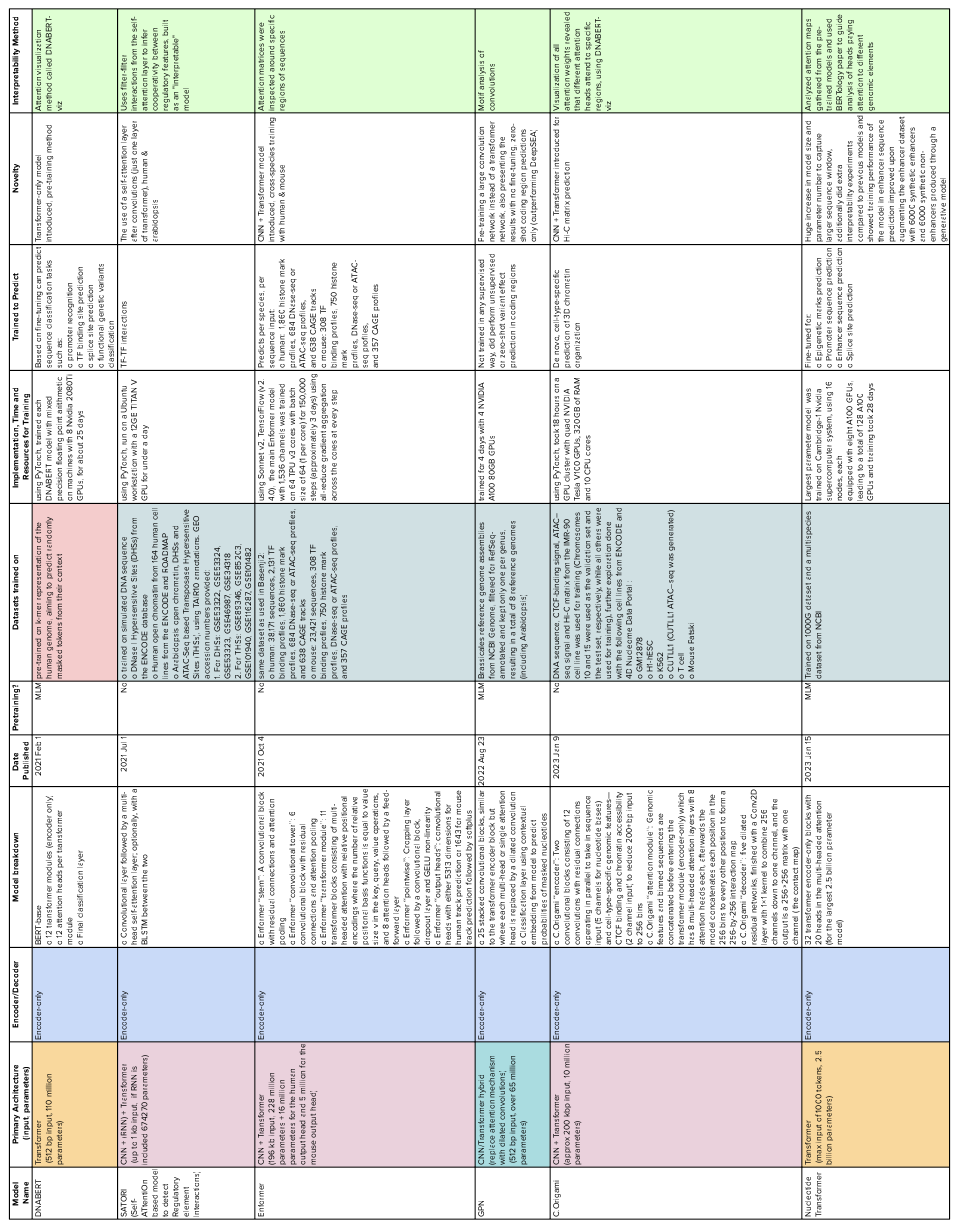}
  \captionof{table}{An in-depth summary of the recently proposed deep learning models covered in this review. In the architecture column the yellow colour indicates a transformer LLM model, the pink colour indicates a transformer-hybrid model, and the turquoise colour indicates a non-transformer LLM.The encoder-decoder column in colour-coded by encoder-only models, decoder-only models and encoder-decoder style models. The datasets trained on column has a red colouring for human-only models and blue for cross-species training. The interpretability column is coloured by presence or absence of model-interpretability analysis.}
  \label{suptab1}
\end{figure}
\addtocounter{table}{-1}%
\begin{figure}[!ht]
  \centering
  \includegraphics[width=\linewidth, totalheight=19cm]{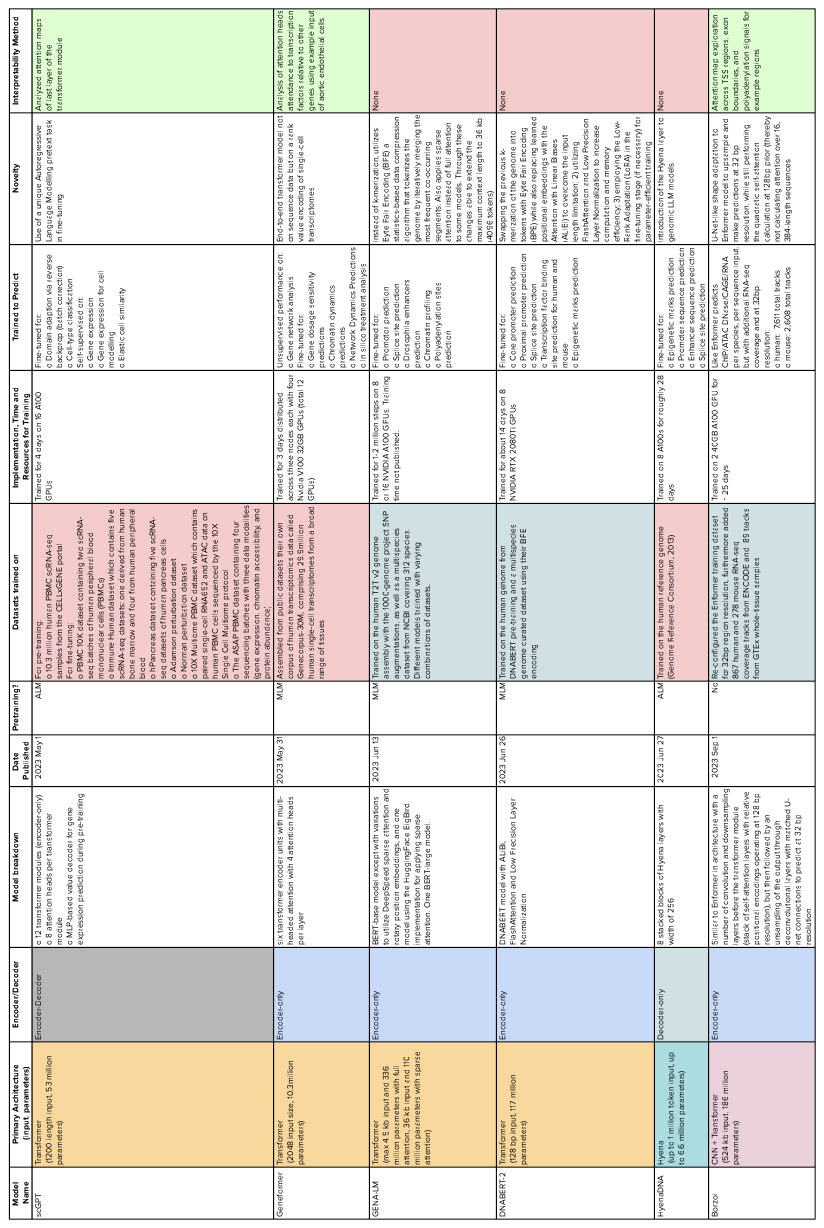}
  \captionof{table}{An in-depth summary of the recently proposed deep learning models covered in this review. In the architecture column the yellow colour indicates a transformer LLM model, the pink colour indicates a transformer-hybrid model, and the turquoise colour indicates a non-transformer LLM. The encoder-decoder column in colour-coded by encoder-only models, decoder-only models and encoder-decoder style models. The datasets trained on column has a red colouring for human-only models and blue for cross-species training. The interpretability column is coloured by presence or absence of model-interpretability analysis.}
\end{figure}

\begin{figure}[!ht]
\centering
\includegraphics[width=\textwidth]{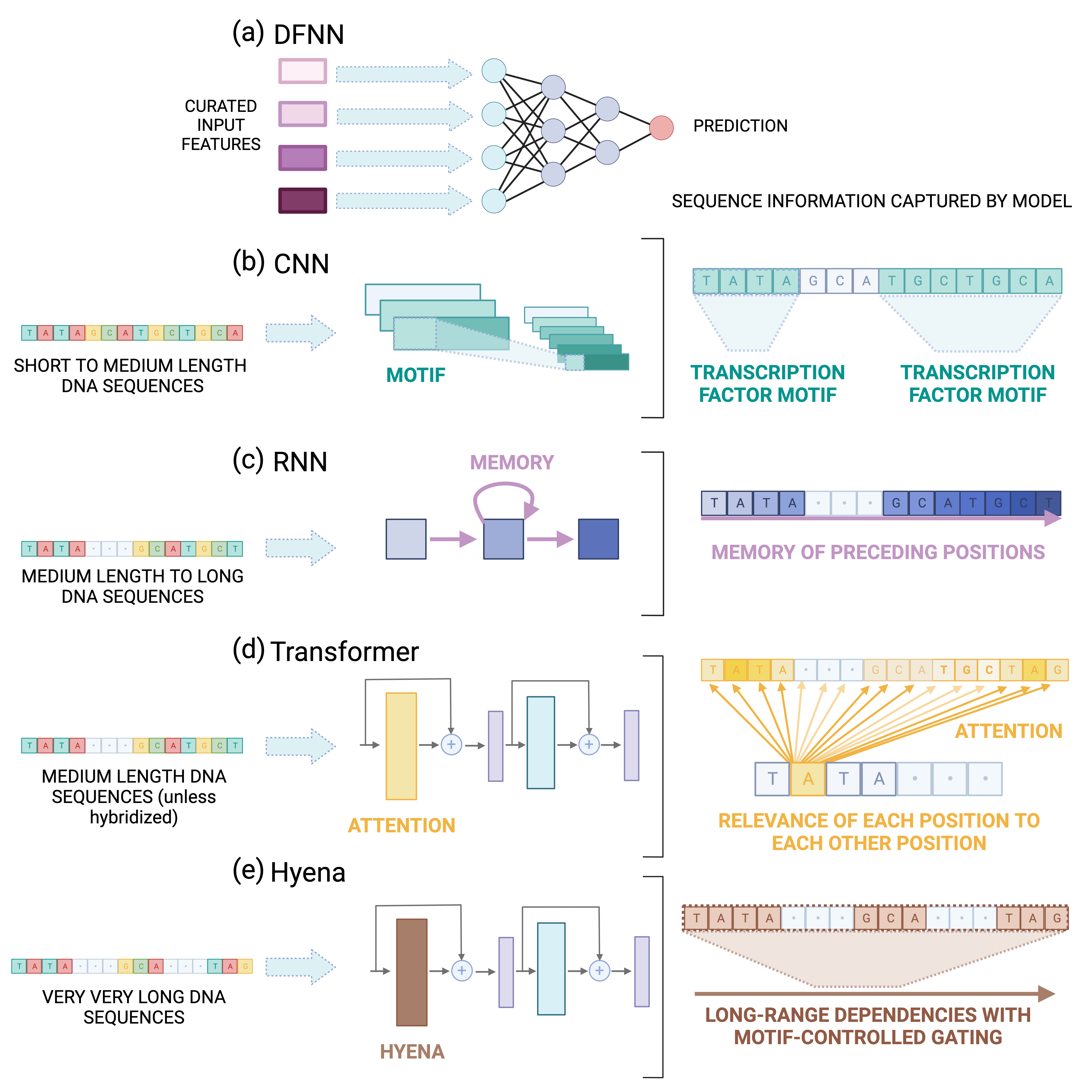}
\caption{\textbf{A comparison of how the strengths of different DL models act on genomic regulatory annotation prediction} All sequence lengths of DNA are given in approximate measures of kilobases (kb) that are typically used as input by models in this review with the specific architecture. (a) A DFNN is capable of taking a curated feature set, either manually curated or taken from the output of another model, and making a prediction. (b) A CNN can directly take as input a DNA sequence, and use convolutions to scan across a sequence to capture local patterns, or \textbf{motifs}, within a DNA sequence. This allows a CNN to pick up motifs that repeat across DNA, like promoters and genes. (c) An RNN can take as input a longer DNA sequence and scan along its entirety while retaining a \textbf{memory} of what it has already seen. The RNN can use its memory as context to inform sequence information it has yet to see. (d) A transformer can take as input a relatively short DNA sequence and use \textbf{attention} to ``attend'' to every position within the inputted sequence. Attention allows the transformer to capture longer-range dependencies and draw relationships between every combination of sequence position in terms of relevant importance to another within the input sequence. (e) A Hyena layer has the same organization as a transformer layer, only replacing the attention mechanism with a Hyena Operator. This Hyena Operator applies a series of implicitly parameterized long convolutions (across the whole sequence) to projections of the input sequence, such that motifs found in earlier convolutions can act as gates to ``remembering'' or ``forgetting'' sequence information in later convolutions, similar to an RNN.}
\label{suppfig1}
\end{figure}

\begin{figure}[!ht]
\centering
\includegraphics[width=\linewidth]{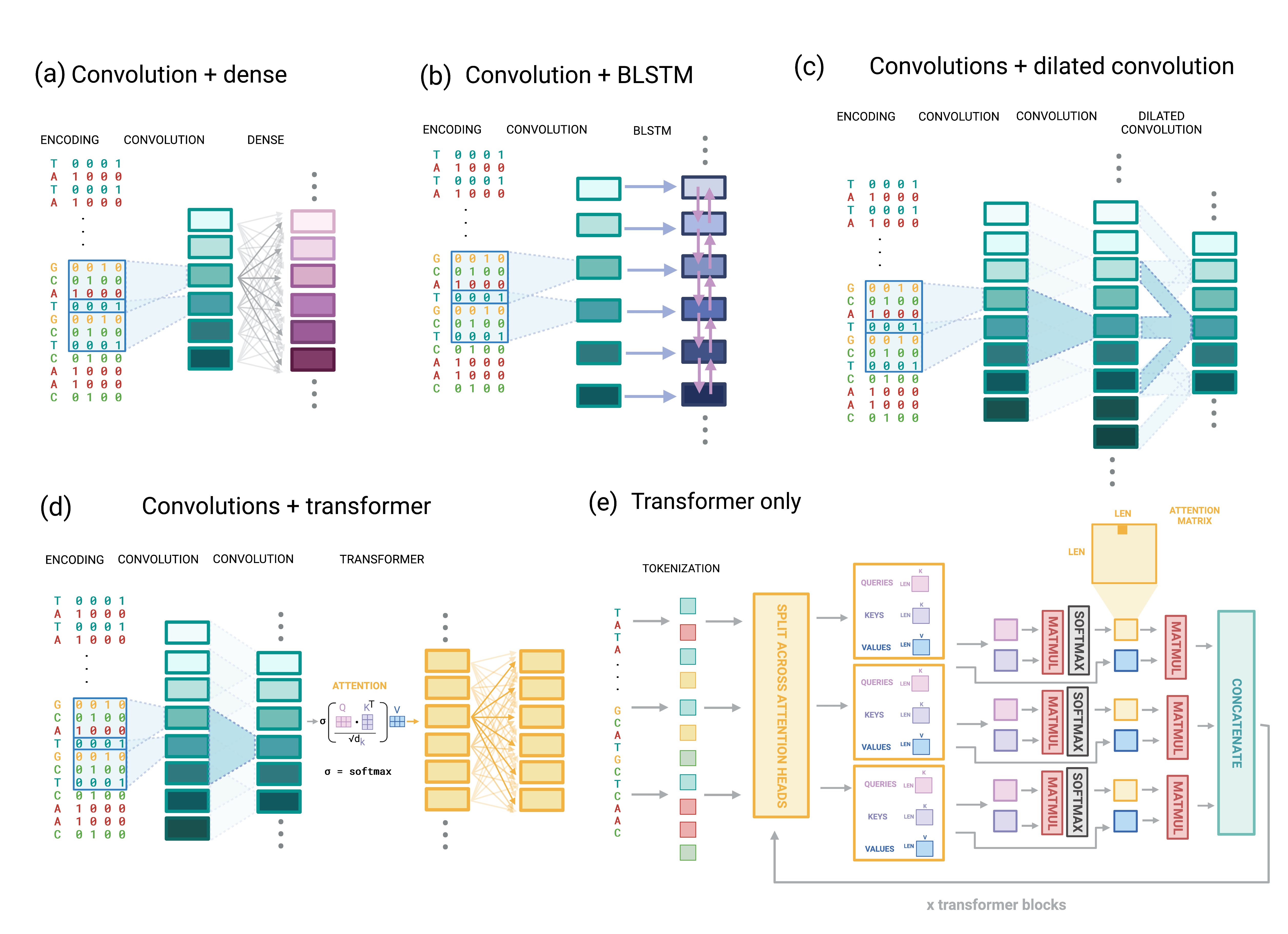}
\caption{\textbf{How different layers types are combined to form deep learning models capable of predicting regulatory annotations} (a) Densely or fully-connected layers share information across all internal nodes, or hidden states. This allows these layers to capture non-linear combinations of relationships across multiple features and find a space to separate and classify them. In the models explored in this paper they are often used after convolutional layers, which track and summarize local information within hidden states. This allows the convolutions to find patterns within sequences like common motifs found in promoters. Convolutional layers followed by dense layers are very common in early architectures for non-coding variant effect prediction, like DeepBind and DeepSEA. (b) Combination of recurrent layers from an RNN, particularly a Bi-directional Long Short Term Memory network (BLSTM), with convolutional layers. These BLSTMs have access to the hidden states of previous convolutional layers, allowing them to keep a memory of previous sequence relationships, and how they relate to information at the current sequence position, as summarized by convolutions. (c) Basenji and models introduced after it, use dilated convolutions to capture a wider range of sequence data by introducing ``holes'' or gaps within kernels at later convolutional layers. (d) The Enformer model uses a transformer module that captures the relevance between preceding convolutional layer hidden states, representing bins of sequences, through an attention calculation. The attention score of all hidden states in relation to one another enforces importance of certain elements and enables long-range modelling. (e) Models like DNABERT can take genomic sequences and tokenize them using a DNA-based vocabulary. These tokenized sequences are then split across multiple heads in multi-headed attention. The keys, queries and values are all derived from the tokenized sequences and the attention matrix sometimes used to report interpretability scores from transformer models is often extracted from the multiplication and softmaxing of the keys and queries. This mechanism allows the attention scores to directly reflect the tokenized sequences as opposed to mapping back to convolutional layers.}
\label{suppfig2}
\end{figure}

\end{document}